\documentclass[a4paper]{amsart}

\usepackage[T1]{fontenc}

\RequirePackage{amsmath}
\RequirePackage{bm}
\RequirePackage{amssymb}
\RequirePackage{upref}
\RequirePackage{amsthm}
\RequirePackage{enumerate}
\RequirePackage{pb-diagram}
\RequirePackage{amsfonts}
\RequirePackage[mathscr]{eucal}
\RequirePackage{verbatim}
\RequirePackage{xr}
\RequirePackage{graphicx}
\usepackage{calc}
\usepackage{xspace}
\RequirePackage{color}
\RequirePackage{ifthen}
\RequirePackage{fancybox}
\RequirePackage{mathrsfs}

\newcommand{\cf}{cf.\@\xspace}
\newcommand{\resp}{resp.\@\xspace}

\newcommand{\al}{\alpha}
\newcommand{\bet}{\beta}
\newcommand{\ga}{\gamma}
\newcommand{\de}{\delta }
\newcommand{\e}{\epsilon}

\newcommand{\f}{\varphi}
\newcommand{\h}{\eta}

\newcommand{\lam}{\lambda}

\newcommand{\s}{\sigma}

\newcommand{\D}{\varDelta}
\newcommand{\F}{\varPhi}
\newcommand{\Lam}{\varLambda}

\newcommand{\so}{{\mc S_0}}
\newcommand{\socc}{{\mc S_0}}

\newcommand{\const}{\tup{const}}

\newcommand{\msp[1]}[1]{\mspace{#1mu}}

\newcommand{\R}[1][n+1]{{\protect\mathbb R}^{#1}}

\newcommand{\Hh}[1][n+1]{{\protect\mathbb H}^{#1}}

\newcommand{\Cc}{{\protect\mathbb C}}

\newcommand{\N}{{\protect\mathbb N}}

\newcommand{\Z}{{\protect\mathbb Z}}
\newcommand{\eR}{\stackrel{\lower1ex \hbox{\rule{6.5pt}{0.5pt}}}{\msp[3]\R[]}}
\newcommand{\eN}{\stackrel{\lower1ex \hbox{\rule{6.5pt}{0.5pt}}}{\msp[1]\N}}
\newcommand{\eO}{\stackrel{\lower1ex \hbox{\rule{6pt}{0.5pt}}}{\msc O}}

\DeclareMathOperator{\tr}{tr}

\newcommand\ra{\rightarrow}

\newcommand\hra{\hookrightarrow}

\newcommand\pa{\partial}
\newcommand\pde[2]{\frac {\partial#1}{\partial#2}}
 
\newcommand{\un}{\infty}
\newcommand{\A}{\forall}

\newcommand{\set}[2]{\{\,#1\colon #2\,\}}
\newcommand{\uu}{\cup}

\newcommand{\uuu}{\bigcup}

\newcommand{\uud}{ \stackrel{\lower 1ex \hbox {.}}{\uu}}
\newcommand{\uuud}[1]{ \stackrel{\lower 1ex \hbox {.}}{\uuu_{#1}}}
\newcommand\su{\subset}

\newcommand{\sminus}[1][28]{\raise 0.#1ex\hbox{$\scriptstyle\setminus$}}

\newcommand{\wed}{\wedge}

\newcommand{\abs}[1]{\lvert#1\rvert}

\newcommand{\norm}[1]{\lVert#1\rVert}

\newcommand{\nnorm}[1]{| \mspace{-2mu} |\mspace{-2mu}|#1| \mspace{-2mu}
|\mspace{-2mu}|}
\newcommand{\spd}[2]{\protect\langle #1,#2\protect\rangle}
\newcommand{\spdd}[2]{\protect\langle\protect\langle #1,#2\protect\rangle\protect\rangle}

\newcommand{\tit}{\textit}

\newcommand{\tup}{\textup}

\newcommand{\mc}{\protect\mathcal}
\newcommand{\msc}{\protect\mathscr}

\providecommand{\bysame}{\makebox[3em]{\hrulefill}\thinspace}

\newcommand{\bt}{\begin{thm}}
\newcommand{\bl}{\begin{lem}}
\newcommand{\bc}{\begin{cor}}
\newcommand{\bd}{\begin{definition}}
\newcommand{\bpp}{\begin{prop}}
\newcommand{\br}{\begin{rem}}
\newcommand{\bn}{\begin{note}}
\newcommand{\be}{\begin{ex}}
\newcommand{\bes}{\begin{exs}}
\newcommand{\bb}{\begin{example}}
\newcommand{\bbs}{\begin{examples}}
\newcommand{\ba}{\begin{axiom}}
\newcommand{\bas}{\begin{assumption}}

\newcommand{\et}{\end{thm}}
\newcommand{\el}{\end{lem}}
\newcommand{\ec}{\end{cor}}
\newcommand{\ed}{\end{definition}}
\newcommand{\epp}{\end{prop}}
\newcommand{\er}{\end{rem}}
\newcommand{\en}{\end{note}}
\newcommand{\ee}{\end{ex}}
\newcommand{\ees}{\end{exs}}
\newcommand{\eb}{\end{example}}
\newcommand{\ebs}{\end{examples}}
\newcommand{\ea}{\end{axiom}}
\newcommand{\eas}{\end{assumption}}

\newcommand{\bp}{\begin{proof}}
\newcommand{\ep}{\end{proof}}
\newcommand{\eps}{\renewcommand{\qed}{}\end{proof}}

\newcommand{\bal}{\begin{align}}

\newcommand{\bi}[1][1.]{\begin{enumerate}[\upshape #1]}
\newcommand{\bia}[1][(1)]{\begin{enumerate}[\upshape #1]}
\newcommand{\bin}[1][1]{\begin{enumerate}[\upshape\bfseries #1]}
\newcommand{\bir}[1][(i)]{\begin{enumerate}[\upshape #1]}
\newcommand{\bic}[1][(i)]{\begin{enumerate}[\upshape\hspace{2\cma}#1]}
\newcommand{\bis}[2][1.]{\begin{enumerate}[\upshape\hspace{#2\parindent}#1]}
\newcommand{\ei}{\end{enumerate}}

\newcommand\ndots{\raise 0.47ex \hbox {,}\hskip0.06em\cdots %
     \raise 0.47ex \hbox {,}\hskip0.06em} 

\newcommand{\q}{\quad}
\newcommand{\qq}{\qquad}

\newcommand\nd{\noindent}

\newcommand{\nt}{\notag}

\newskip\Csmallskipamount                                                
\Csmallskipamount=\smallskipamount
\newskip\Cmedskipamount
\Cmedskipamount=\medskipamount
\newskip\Cbigskipamount
\Cbigskipamount=\bigskipamount

\newcommand\cvs{\vspace\Csmallskipamount}   
\newcommand\cvm{\vspace\Cmedskipamount}

\newskip\csa
\csa=\smallskipamount

\newskip\cma
\cma=\medskipamount

\newskip\cba
\cba=\bigskipamount

\newdimen\spt
\newcommand\citem{\cvs\advance\itemno by
1{(\romannumeral\the\itemno})\hskip3pt}
\newcommand{\bitem}{\cvm\nd\advance\itemno by
1{\bf\the\itemno}\hspace{\cma}}

\newcount\itemno
\newcommand{\las}[1]{\label{S:#1}}

\newcommand{\lae}[1]{\label{E:#1}}
\newcommand{\lat}[1]{\label{T:#1}}
\newcommand{\lal}[1]{\label{L:#1}}
\newcommand{\lad}[1]{\label{D:#1}}
\newcommand{\lac}[1]{\label{C:#1}}

\newcommand{\lar}[1]{\label{R:#1}}

\newcommand{\rs}[1]{Section~\ref{S:#1}}

\newcommand{\rt}[1]{Theorem~\ref{T:#1}}
\newcommand{\rl}[1]{Lemma~\ref{L:#1}}
\newcommand{\rd}[1]{Definition~\ref{D:#1}}

\newcommand{\rr}[1]{Remark~\ref{R:#1}}

\newcommand{\re}[1]{\eqref{E:#1}}
\newcommand{\frc}[1]{Corollary~\ref{C:#1} on page~\tup{\pageref{C:#1}}}
\newcommand{\frt}[1]{Theorem~\ref{T:#1} on page~\tup{\pageref{T:#1}}}
\newcommand{\frl}[1]{Lemma~\ref{L:#1} on page~\tup{\pageref{L:#1}}}

\newcommand{\fre}[1]{\eqref{E:#1} on page~\tup{\pageref{E:#1}}}

\newcommand{\frs}[1]{Section~\ref{S:#1} on page~\tup{\pageref{S:#1}}}

\newskip\thmskip
\thmskip=\parindent

\newskip\hsk
\newenvironment{hinw}{\labelsep=0pt\begin{list}{}{\labelsep=0pt\itemindent=0pt\labelwidth=0pt\leftmargin=\parindent\rightmargin=0pt\partopsep=\cba}%
\item\it\nopagebreak\nopagebreak}%
{\end{list}}

\newcommand\bh{\begin{hinw}}
\newcommand{\eh}{\end{hinw}}

\newtheoremstyle{normal}
  {\cba}
  {\cba}
  {}
  {\thmskip}
  {\bfseries}
  {.}
  {\hsk}
  {}

\newtheoremstyle{abschnitt}
  {\cba}
  {\cba}
  {}
  {\thmskip}
  {\bfseries}
  {.}
  {\hsk}
  {}

\newtheoremstyle{italic}
  {\cba}
  {\cba}
  {\itshape}
  {\thmskip}
  {\bfseries}
  {.}
  {\hsk}
  {}

\newtheoremstyle{aufgaben}
  {\cba}
  {\cba}
  {}
  {}
  {\normalsize\bfseries}
  {.}
  {\hsk}
  {}

\newtheoremstyle{break}
  {\cba}
  {\cba}
  {\itshape}
  {}
  {\bfseries}
  {.}
  {\newline}
  {}

\theoremstyle{italic}
\newtheorem{thm}[subsection]{Theorem}
\newtheorem{lem}[subsection]{Lemma}
\newtheorem{prop}[subsection]{Proposition}
\newtheorem{cor}[subsection]{Corollary}

\theoremstyle{normal}
\newtheorem{rem}[subsection]{Remark}
\newtheorem{definition}[subsection]{Definition}
\newtheorem{example}[subsection]{Example}
\newtheorem{examples}[subsection]{Examples}
\newtheorem{ex}[subsection]{Exercise}
\newtheorem{note}[subsection]{}
\newtheorem{axiom}[subsection]{Axiom}
\newtheorem{assumption}[subsection]{Assumption}

\theoremstyle{aufgaben}
\newtheorem{exs}[subsection]{Exercises}

\numberwithin{equation}{section}
\numberwithin{figure}{section}

\newenvironment{textequation}[1][0.8]
{\begin{equation}
\begin{aligned}
\begin{minipage}{#1\linewidth}}
{\end{minipage}
\end{aligned}
\end{equation}
\ignorespacesafterend}

\newcommand{\btext}{\begin{textequation}}
\newcommand{\etext}{\end{textequation}}

\def\hinweis{\@startsection{subsection}{2}%
 \z@{0.7\linespacing\@plus 0.5\linespacing}{0.7\linespacing}%
{\normalfont\itshape\indent}}

\newcounter{hours}\newcounter{minutes}
\newcommand{\printtime}{%
\setcounter{hours}{\time/60}%
\setcounter{minutes}{\time-\value{hours}*60}%
\ifthenelse{\value{minutes}<10}{\thehours :0\theminutes}{\thehours:\theminutes}}

\usepackage[german,english]{babel}
\usepackage{graphicx}
\RequirePackage{amsmath}
\RequirePackage{bm}
\RequirePackage{amssymb}
\RequirePackage{upref}
\RequirePackage{amsthm}
\RequirePackage{enumerate}
\RequirePackage{pb-diagram}
\RequirePackage{amsfonts}
\RequirePackage[mathscr]{eucal}
\RequirePackage{verbatim}
\RequirePackage{xr}
\RequirePackage{graphicx}
\RequirePackage{mathrsfs}
\usepackage{calc}
\usepackage{xspace}
\usepackage{dsfont}

\makeatletter
\RequirePackage{color}
\newcommand{\ann}[1]{\renewcommand{\@makefnmark}{\mbox{$^{\color{red}{\@thefnmark}}$}}%
\footnote {#1}}
\makeatother

\RequirePackage{upref}
\RequirePackage{amsthm}
\RequirePackage{enumerate}
\usepackage[mathscr]{eucal}

\usepackage{xr-hyper}

\usepackage{calc}

\newlength{\oddsidemarginlength}
\newlength{\topmarginlength}

\newcounter{numberoflines}
\newcounter{tempcc}
\oddsidemargin=\oddsidemarginlength
\evensidemargin=\oddsidemargin
\topmargin=\topmarginlength
\usepackage[colorlinks=true,linkcolor=blue,citecolor=blue,urlcolor=blue,hypertexnames=false]{hyperref}  

\begin{document}

\flushbottom


\title[Extending solutions of quantum gravity past the singularity]{Extending the solutions and the equations of quantum gravity past the big bang singularity}

\author{Claus Gerhardt}
\address{Ruprecht-Karls-Universit\"at, Institute for Mathematics,
Im Neuenheimer Feld 205, 69120 Heidelberg, Germany}
\email{\href{mailto:gerhardt@math.uni-heidelberg.de}{gerhardt@math.uni-heidelberg.de}}
\urladdr{\href{http://www.math.uni-heidelberg.de/studinfo/gerhardt/}{https://www.math.uni-heidelberg.de/studinfo/gerhardt/}}

%


\date{\today}
%


\begin{abstract} 
In \cite{cg:qgravity-book2} we recently proved that in our model of quantum gravity the solutions to the quantized version of the full Einstein equations or to the Wheeler-DeWitt equation could be expressed as products of spatial and temporal eigenfunctions, or eigendistributions, of self-adjoint operators acting in corresponding separable Hilbert spaces. Moreover, near the big bang singularity we derived sharp asymptotic estimates for the temporal eigenfunctions. In this paper we show that, by using these estimates, there exists a complete sequence of unitarily equivalent eigenfunctions which can be extended past the singularity by even or odd mirroring as sufficiently smooth functions such that the extended functions are solutions of the appropriately extended  equations valid in $\R[]$ in the classical sense. We also use this phenomenon to explain the missing antimatter.
\end{abstract}

\maketitle

\tableofcontents

\setcounter{section}{0}

\section{Introduction}\las{1} 
A unified quantum theory incorporating the four fundamental forces of nature is one of the major open problems in physics. The Standard Model combines electromagnetism, the strong force and the weak force, but ignores gravity. The quantization of gravity is therefore a necessary first step to achieve a unified quantum theory.      
 
General relativity is a Lagrangian theory, i.e., the Einstein equations are derived as the Euler-Lagrange equation of the Einstein-Hilbert functional 
\begin{equation}
\int_N(\bar R-2\Lam),
\end{equation}
where $N=N^{n+1}$, $n\ge 3$, is a globally hyperbolic Lorentzian manifold, $\bar R$ the scalar curvature and $\Lam$ a cosmological constant. We also omitted the integration density in the integral. In order to apply a Hamiltonian description of general relativity, one usually defines a time function $x^0$ and considers the foliation of $N$ given by the slices
\begin{equation}
M(t)=\{x^0=t\}.
\end{equation}
We may, without loss of generality, assume that the spacetime metric splits
\begin{equation}\lae{1.3}
d\bar s^2=-w^2(dx^0)^2+g_{ij}(x^0,x)dx^idx^j,
\end{equation}
\cf \cite[Theorem 3.2]{cg:qgravity}. Then, the Einstein equations also split into a tangential part
\begin{equation}
G_{ij}+\Lam g_{ij}=0
\end{equation}
and a normal part
\begin{equation}
G_{\al\bet}\nu^\al\nu^\bet-\Lam=0,
\end{equation}
where the naming refers to the given foliation. For the tangential Einstein equations one can define equivalent Hamilton equations due to the groundbreaking paper by Arnowitt, Deser and Misner \cite{adm:old}. The normal Einstein equations can be expressed by the so-called Hamilton condition
\begin{equation}\lae{1.6}
\mc H=0,
\end{equation}
where $\mc H$ is the Hamiltonian used in defining the Hamilton equations. In the canonical quantization of gravity the Hamiltonian is transformed  to a partial differential operator of hyperbolic type $\hat{\mc H}$ and the possible quantum solutions of gravity are supposed to satisfy the so-called Wheeler-DeWitt equation
\begin{equation}\lae{1.7}
\hat{\mc H}u=0
\end{equation}
in an appropriate setting, i.e., only the Hamilton condition \re{1.6} has been quantized, or equivalently, the normal Einstein equation, while the tangential Einstein equations have been ignored.

In \cite{cg:qgravity} we solved the equation \re{1.7} in a fiber bundle $E$ with base space $\socc$,
\begin{equation}
\socc=\{x^0=0\}\equiv M(0),
\end{equation}
and fibers $F(x)$, $x\in\socc$,
\begin{equation}
F(x)\su T^{0,2}_x(\socc),
\end{equation}
the elements of which are the positive definite symmetric tensors of order two, the Riemannian metrics in $\socc$. The hyperbolic operator $\hat{\mc H}$ is then expressed in the form 
\begin{equation}\lae{1.10}
\hat{\mc H}=-\D-(R-2\Lam)\f,
\end{equation}
where $\D$ is the Laplacian of the DeWitt metric given in the fibers, $R$ the scalar curvature of the metrics $g_{ij}(x)\in F(x)$, and $\f$ is defined by
\begin{equation}\lae{1.11}
\f^2=\frac{\det g_{ij}}{\det\rho_{ij}},
\end{equation}
where $\rho_{ij}$ is a fixed metric in $\so$ such that instead of densities we are considering functions. 

The Wheeler-DeWitt equation only represents the quantization of the normal Einstein equations and ignores the tangential Einstein equations. In order to quantize the full Einstein equations we incorporated the Hamilton condition into the right-hand side of the Hamilton equations to obtain a scalar evolution equation such that the Hamilton equations and this scalar evolution equation are equivalent to the full Einstein equations, \cf \cite[Theorem 1.3.4, p. 12]{cg:qgravity-book2}.  For the quantization of this evolution equation we defined the base space of the fiber bundle $E$ to be the Cauchy hypersurface $(\socc,\bar\s_{ij})$ of the quantized spacetime, where $\bar\s_{ij}$ is the induced metric. We also choose the metric $\rho_{ij}$ in \re{1.11} to be equal to $\bar\s_{ij}$. The result of this quantization was a hyperbolic equation in $E$.

The fibers $F(x)$ over $x\in\so$ are Riemannian metrics $g_{ij}(x)$ if external fields are excluded. In an appropriate local trivialization we obtained a coordinate system $(\xi^a)$, $0\le a \le m$, 
\begin{equation}\nt
m=\frac{(n-1)(n+2)}2,
\end{equation}
$n=\dim\so$, such that the metrics $g_{ij}$ can be written
\begin{equation}\nt
g_{ij}=t^\frac4n \s_{ij},
\end{equation}
where
\begin{equation}\nt
0<t=\xi^0<\un
\end{equation}
and the metric $\s_{ij}$ belongs to the hypersurface or subbundle
\begin{equation}\nt
M=\{t=1\}\su E.
\end{equation}
The solutions $u$ then depend on the variables $(t,\s_{ij},x)$, where $\s_{ij}$ does not depend on $t$ and $t$ not on $x$. We refer to $t$ as quantum time and $x,\s_{ij}$ as spatial variables.

In the papers \cite{cg:qgravity3,cg:qgravity4} we   could express $u$ as a product of eigenfunctions
\begin{equation}
u=w\hat v v,
\end{equation}
where $w=w(t)$ is the temporal eigenfunction, $\hat v=\hat v(\s_{ij}(x))$ can be identified with an eigenfunction of the Laplacian of the symmetric space
\begin{equation}
X=SL(n,\R[])/SO(n)
\end{equation}
such that
\begin{equation}
\hat v(\bar\s_{ij}(x))=1\qq\A\, x\in\so,
\end{equation}
where $\bar\s_{ij}$ is the fixed induced metric of $\so$.  The eigenfunctions $\hat v$ represent the elementary gravitons  corresponding to the degrees of freedom in choosing the entries of  Riemannian metrics with determinants equal to one. These are all the degrees of freedom available because of the coordinate system invariance: For any smooth Riemannian metric there exists an atlas such that the determinant of the metric is equal to one, \cf \cite[Lemma 3.2.1, p. 74]{cg:qgravity-book2}. The function $v$ is an eigenfunction of an essentially self-adjoint differential operator in $\so$.

At first, the temporal eigenfunctions $w$ were only the solutions of an ODE. Later, in \cite[Section 5]{cg:qgravity4} we proved that they were the eigenfunctions of an essentially self-adjoint differential operator in $\R[]_+$, provided $n$ is sufficiently large and $\Lam<0$ and the Cauchy hypersurface $(\socc,\bar\s_{ij})$ is either a space of constant curvature like $\R[n]$ and $\Hh[n]$ or a metric product of the form
\begin{equation}\lae{1.15}
\so=\R[n_1]\times M_0,
\end{equation}
where $M_0$ is a smooth, compact and connected manifold of dimension $n-n_1$,
\begin{equation}
\dim M_0=n-n_1=n_0,
\end{equation}
and where
\begin{equation}
\bar\s=\de\otimes g
\end{equation}
is a metric product; $\de$ is the  standard Euclidean metric and $g$ a Riemannian metric in $M_0$, \cf \cite[Section 5]{cg:qgravity4}.

 But in \cite[Chapter 4.2]{cg:qgravity-book2} we were able to prove this property for arbitrary $n\ge 3$ and $\Lam<0$ and, in case $n=3$, even for $\Lam >0$ by introducing an additional  scalar fields map in the action functional, i.e., a map 
\begin{equation}
\F:N\ra \R[k],
\end{equation}
where $N=I\times \so$ is the original spacetime which is to be quantized. Let $(\bar g_{\al\bet})$ be the Lorentzian metric in $N$, the scalar field Lagrangian is defined by
\begin{equation}\lae{2.4.1.2.4}
L_S=-\frac12 \bar g^{\al\bet}\ga_{ab}\F^a_\al\F^b_\bet\sqrt{\abs{\bar g}},
\end{equation}
i.e., without a zero order term, $(\ga_{ab})$ is the Euclidean metric in $\R[k]$.

The temporal eigenfunctions $w$ then have to satisfy the ODE
\begin{equation}\lae{1.20}
\begin{aligned}
&\frac n{16(n-1)} t^{-(m+k)}\frac\pa{\pa t}\big (t^{(m+k)} \pde wt\big )+t^{-2}(\abs\lam^2+\rho^2-\frac12\abs{\theta_0}^2)w\\
&\qq + t^{2-\frac4n}\{(n-1)\abs\xi^2+\bar\mu_l\}w+(n-2) t^2\Lam w=0
\end{aligned}
\end{equation}
in $0<t<\un$, where
\begin{equation}
\abs\lam^2+\rho^2
\end{equation}
is an eigenvalue of an elementary graviton,
\begin{equation}
\abs{\theta_0}^2,
\end{equation}
an eigenvalue of $-\D_{\R[k]}$ and
\begin{equation}
(n-1)\abs\xi^2+\bar\mu_l
\end{equation}
with $\xi\in \R[{n_1}]$ an eigenvalue of the spatial self-adjoint operator acting in \re{1.15}.

Using the abbreviations
\begin{equation}\lae{1.24}
\mu_0=\frac{16(n-1)}n(\abs \lam^2+\abs\rho^2-\frac12 \abs{\theta_0}^2),
\end{equation}
\begin{equation}\lae{1.25}
m_1=\frac{16(n-1)}n\{(n-1)\abs\xi^2+\bar\mu_l\}
\end{equation}
and
\begin{equation}\lae{2.4.2.68.4}
m_2=\frac{16(n-1)(n-2)}n 
\end{equation}
we can rewrite the equation \re{1.20} in the form
\begin{equation}\lae{1.27}
\begin{aligned}
t^{-(m+k)}\frac\pa{\pa t}\big (t^{(m+k)} \pde wt\big )+t^{-2}\mu_0w
+ t^{2-\frac4n} m_1 w+ t^2 m_2 \Lam w=0.
\end{aligned}
\end{equation}
This equation can be treated as an eigenvalue equation provided
\begin{equation}\lae{1.28}
\bar\mu=\mu_0-\frac {(m+k-1)^2}4<0.
\end{equation}
Let us recall that
\begin{equation}
m=\frac{(n-1)(n+2)}2.
\end{equation}
and
\begin{equation}\lae{1.30}
\rho^2=\frac{(n-1)^2n}{12}.
\end{equation}
There are two ways how to treat \re{1.27} as an eigenvalue equation: First, the cosmological constant $\Lam$, or better $-\Lam$ can be looked at as an implicit eigenvalue, or secondly, if we consider $\Lam<0$ to be fixed, we could try to solve the eigenvalue problem
\begin{equation}\lae{1.31}
-t^{-(m+k)}\frac\pa{\pa t}\big (t^{(m+k)} \pde wt\big )-t^{-2}\mu_0w
- t^2 m_2 \Lam w=\lam t^{2-\frac4n} w
\end{equation}
in $(0,\un)$, where $\lam>0$ is a yet unknown eigenvalue such that $\lam$ would be equal to the spatial eigenvalue, i.e.,
\begin{equation}
\lam=m_1=\frac{16(n-1)}n\{(n-1)\abs\xi^2+\bar\mu_l\}.
\end{equation} 
In this case the corresponding eigenfunction $w$ would be a solution of \re{1.27}, i.e., it would be a temporal eigenfunction of our model of quantum gravity. We solved the implicit as well as the explicit eigenvalue problem in \cite[Chapter 4]{cg:qgravity-book2} by choosing $k$ in \re{1.28} sufficiently large such that $\bar\mu<0$. 

Since $\mu_0$ is in general positive, unless we choose $\abs {\theta_0}$ large which is not always possible or desirable, we considered the orthogonally equivalent function
\begin{equation}\lae{1.33}
u=t^{\frac{m+k-1}2}w
\end{equation}
which satisfies the equation
\begin{equation}\lae{1.34}
-t^{-1}\frac\pa{\pa t}\big (t \pde ut\big )-t^{-2}\bar\mu u
+t^2 m_2^2  u=\lam t^{2-\frac4n}u,
\end{equation}
where
\begin{equation}
\bar\mu=\mu_0-{\bigg(\frac{m+k-1}2\bigg)}^2
\end{equation}
which is negative if $k\in\N$ is large enough.

In \cite[Theorem 3.4.9, p. 86]{cg:qgravity-book2} we proved
\bt\lat{1.1}
Let  $u\in \mc H_2$ satisfy the equation \re{1.34} which we express in the form
\begin{equation}\lae{1.36}
A_1u=-t^{-1}\frac\pa{\pa t}\big (t \pde ut\big )+t^{-2}\mu^2 u
+t^2 m_2^2  u=\lam t^{2-\frac4n} u,
\end{equation}
where the constants $\mu, m_2$ and $\lam$ are strictly positive. Since $\mu$ is especially important, let us emphasize that
\begin{equation}
\mu^2=-\bar\mu=\frac{(m+k-1)^2}4-\mu_0
\end{equation}
and $\mu_0>0$. Then, there exists $0<t_0<1$ and positive constants $p,c_1,c_2$ such that $u$ does not vanish in the interval $(0,t_0]$ and can be estimates by
\begin{equation}\lae{1.38}
c_1 t^p\le \abs {u(t)}\le c_2 t^\mu\qq\A\,t\in (0,t_0],
\end{equation}
where $p$,
\begin{equation}\lae{1.39}
\mu<p<\frac{m+k-1}2,
\end{equation}
is arbitrary but fixed.
\et
Here, we adapted the wording slightly to reflect the present assumptions, \cf \cite[Theorem 4.2.4, p. 118]{cg:qgravity-book2}.

If we combine gravity with the forces of the Standard Model then we cannot quantize the full Einstein equations but only the normal Einstein equation, i.e., the Hamilton condition. As a result we obtain the Wheeler-DeWitt equation which again can be solved by a product of spatial and temporal eigenfunctions or eigendistributions. In this case the temporal eigenfunction equation has the form, after using the same ansatz as before,
\begin{equation}\lae{1.40}
-t^{-1}\frac\pa{\pa t}\big (t \pde ut\big )-t^{-2}\bar\mu u
+t^2 m_2^2  u=\lam t^{-\frac23}u,
\end{equation}
where
\begin{equation}
\bar\mu=\mu_0-{\bigg(\frac{m+k-1}2\bigg)}^2.
\end{equation}
Comparing this equation with equation \re{1.34} there are two differences: First, the term $\mu_0$ does not depend on $\abs{\theta_0}$ 
\begin{equation}
\mu_0=\frac{16(n-1)}n(\abs \lam^2+\abs\rho^2)
\end{equation}
since we had to choose $\theta_0=0$, and secondly, the exponent of $t$ on the right-side is $-\frac23$. The first difference implies that only by requiring $k$ to be large we could enforce $\bar\mu<0$ and the negative exponent that the estimate \re{1.38} is slightly worse, but still good enough for our purpose. Indeed, we proved in \cite[Theorem 5.5.5, p. 145]{cg:qgravity-book2}
\bt\lat{1.2}
Let  $u\in \mc H_2$ satisfy the equation 
\begin{equation}\lae{1.43}
A_1u=-t^{-1}\frac\pa{\pa t}\big (t \pde ut\big )+t^{-2}\mu^2 u
+t^2 m_2^2  u=\lam t^{-\frac23} u,
\end{equation}
where the constants $\mu, m_2$ and $\lam$ are strictly positive. Since $\mu$ is especially important, let us emphasize that
\begin{equation}\lae{1.44}
\mu^2=-\bar\mu=\frac{(m+k-1)^2}4-\mu_0
\end{equation}
and $\mu_0>0$. Then, for any small $\e_0>0$,  there exist $0<t_0<1$ and positive constants $p,c_1,c_2$ such that $u$ does not vanish in the interval $(0,t_0]$ and can be estimated by
\begin{equation}\lae{1.45}
c_1 t^p\le \abs {u(t)}\le c_2 t^{\mu-\e_0}\qq\A\,t\in (0,t_0],
\end{equation}
where $p$,
\begin{equation}\lae{1.46}
\mu<p<\frac{m+k-1}2,
\end{equation}
is arbitrary but fixed.
\et
The eigenvalue equations \re{1.36} and  \re{1.43} in the Hilbert space $\mc H_2$ can both be solved by complete sequences of mutually orthogonal eigenfunctions $u_i$ with corresponding positive eigenvalues $\lam_i$ of multiplicity one satisfying 
\begin{equation}
0<\lam_0<\lam_1<\lam_2<\cdots
\end{equation}
and
\begin{equation}
\lim_{i\ra \un}\lam_i =\un.
\end{equation}
For a proof see \cite[Theorem 3.4.5, p. 84]{cg:qgravity-book2} and \frt{3.8}, where we shall prove a corresponding result for a more general right-hand side which includes our two cases.

As a corollary, which we like to  formulate as a theorem, we deduce:
\bt\lat{1.3}
Let  $w_i\in \hat{\mc H}_2$ be related to a function $u_i$ by
\begin{equation}
w_i=t^{-\frac{m+k-1}2}u_i
\end{equation}
 and assume that $u_i\in \mc H_2$ satisfies an equation of the form 
\begin{equation}\lae{1.50}
A_1u=-t^{-1}\frac\pa{\pa t}\big (t \pde ut\big )+t^{-2}\mu^2 u
+t^2 m_2^2  u=\lam t^{-\frac23} u,
\end{equation}
where the constants $\mu, m_2$ and $\lam$ are strictly positive and  $\mu$ is defined by 
\begin{equation}
\mu^2=-\bar\mu=\frac{(m+k-1)^2}4-\mu_0
\end{equation}
and $\mu_0>0$. Then, for any small $\e_0>0$ there exists $0<t_0<1$ and positive constants $p,c_1,c_2$,  such that $w_i$ does not vanish in the interval $(0,t_0]$ and can be estimates by
\begin{equation}\lae{1.52}
c_1 t^{p-\frac{m+k-1}2}\le \abs {w_i(t)}\le c_2 t^{\mu-\e_0-\frac{m+k-1}2}\qq\A\,t\in (0,t_0],
\end{equation}
where $p$,
\begin{equation}\lae{1.53}
\mu<p<\frac{m+k-1}2,
\end{equation}
is arbitrary but fixed. Hence, we conclude 
\begin{equation}
\lim_{t\ra 0}\abs{w_i(t)}=\un.
\end{equation}
\et

The eigenfunctions $w_i$ in the previous theorem are the solutions of the original temporal eigenfunctions equation and they  are the eigenfunctions of a self-adjoint operator in a Hilbert space. The $u_i$ are the unitarily equivalent eigenfunctions of a unitarily equivalent self-adjoint operator. In \frs{3} we shall show that the unitarily equivalent eigenfunctions
\begin{equation}
\tilde u_i=t^\frac12 u_i
\end{equation}
can be extended past the singularity by an even reflection as sufficiently smooth functions provided the coefficient $\mu^2$ in \re{1.44} is large enough. More precisely, we shall prove:
\bt\lat{1.4}
 Let  $2\le m_0\in\N$ be arbitrary and assume
 \begin{equation}
\mu+\frac12>m_0,
\end{equation}
  then  
 \begin{equation}\lae{1.57}
\tilde u_i\in C^{m_0}([0,t_0])\q\wed\q \tilde u_i^{(m_0)}(0)=0=\lim_{t\ra 0}\tilde u_i^{(m_0)}(t)
\end{equation}
as well as
\begin{equation}\lae{1.58}
\lim_{t\ra 0}\frac{\tilde u_i^{(k)}(t)}{t^{m_0-k}}=0\qq\A\, 1\le k\le m_0, \,k\in \N,
\end{equation}
where $\tilde u_i^{(k)}$ denotes the $k$-th derivative of $\tilde u_i$. These properties are also valid for the extended functions. 
 \et
Furthermore, we shall prove
\bc
If the assumption of the preceding theorem is satisfied then the extended solutions $\tilde u_i$ also satisfy the extended equations
\begin{equation}\lae{1.59}
-\Ddot{ \tilde u}_i  + t^{-2} {\tilde \mu}^2 \tilde u_i +t^2 m_2^2  \tilde u_i=\lam_i \abs t^q  \tilde u_i
\end{equation}
in $\R[]$, where we have to replace $t^q$ by $\abs t^q$ for obvious reasons. Let us emphasize that the lower order coefficients of the ODE exhibit a singularity in $t=0$ but that both sides of the equation are continuous in the interval $(-\un,\un)$ and vanish in $t=0$.
\ec
Here, the exponent $q$ is any real number satisfying
\begin{equation}
-2<q<2.
\end{equation}

In \rs{5} we shall also use these results to explain the missing antimatter.
\section{The equations of quantum gravity}
The tangential  Einstein equations are equivalent to the Hamilton equations and the normal Einstein equation is equivalent to the Hamilton condition.  By quantizing the Hamilton condition we obtain the Wheeler-DeWitt equation while ignoring the tangential Einstein equations. In order to quantize the full Einstein equations we consider the second Hamilton equations
\begin{equation}
\dot \pi^{ij}=-\frac{\de H}{\de g_{ij}},
\end{equation}
where 
\begin{equation}
H=H_0+H_1
\end{equation}
is the combined Hamilton function of the gravitational Hamiltonian $H_0$ and the scalar field map Hamiltonian $H_1$.
Thus, we infer
\begin{equation}\lae{2.3}
g_{ij}\dot\pi^{ij}=-g_{ij}\frac{\de H}{\de g_{ij}}=-g_{ij}\frac{\de (H_0+H_1)}{\de g_{ij}}.
\end{equation}
On the right-hand side of this evolution equation we then implement the Hamilton condition $H=0$ in the form
\begin{equation} 
p H=0,
\end{equation}
where $0\not=p\in \R[]$ is an arbitrary real number to be determined later. After the quantization of the modified evolution equation \re{2.3} we obtain the hyperbolic equation 
\begin{equation}\lae{4.2.47.4}
\begin{aligned}
&(\frac n2-2-p)\{-\frac n{16(n-1)}t^{-(m+k)}\frac\pa {\pa t}(t^{(m+k)}\dot u)\\
&\qq +t^{-2}\D_Mu+\frac12 t^{-2}\D_{\R[k]}u\}-(n-1)t^{2-\frac4n}\tilde\D_\s u\\
&\qq-pt^{2-\frac4n}R_\s u +2p\Lam u+t^{-2}\D_{\R[k]}u+pC_1u=0.
\end{aligned}
\end{equation}
The preceding equation is evaluated at $(x,t,\s_{ij},\theta^a)$, where $x\in \so$, $t\in \R[]_+$, $\s_{ij}\in M$ is the induced metric of a Cauchy hypersurface of the quantized globally hyperbolic spacetime and $\theta =\theta(x)$ is a coordinate in the fiber $\R[k]$. Let us recall that after quantization the components $\F^a$ of the scalar field are equal to the coordinates $\theta^a$ in $\R[k]$  such that  
\begin{equation}
\F^a(x)=\theta^a(x)\qq\A\, x \in\so
\end{equation}
and
\begin{equation}\lae{4.2.49.4}
C_1=\frac12 t^{2-\frac4n}\s^{ij}\ga_{ab}\theta^a_i\theta^b_j.
\end{equation}
Since we only introduced the scalar field  in order to prove that the temporal "eigenfunctions" are indeed eigenfunctions of  a self-adjoint operator with a pure point spectrum we can simplify the left-hand side of \re{4.2.47.4} by choosing
\begin{equation}
\theta^a(x)=1\qq\A\, x \in\so, \;\A\, 1\le a\le k.
\end{equation}
Hence, we have to solve the equation
\begin{equation}\lae{4.2.51.4}
\begin{aligned}
&(\frac n2-2-p)\{-\frac n{16(n-1)}t^{-(m+k)}\frac\pa {\pa t}(t^{(m+k)}\dot u)\\
&\qq +t^{-2}\D_Mu+\frac12 t^{-2}\D_{\R[k]}u\}-(n-1)t^{2-\frac4n}\tilde\D_\s u\\
&\qq-pt^{2-\frac4n}R_\s u +2pt^2\Lam u+t^{-2}\D_{\R[k]}u=0,
\end{aligned}
\end{equation}
where $u$ depends on $(x,t,\s_{ij},\theta^a)$. The parameter $p\in\R[]$, $p\not=0$, is not yet specified.

As  mentioned before the solution $u$ should be a product of spatial and temporal eigenfunctions. In order to ensure that the temporal eigenfunctions are eigenfunctions of a self-adjoint operator we have to distinguish three cases:

\tit{Case 1}: $\Lam<0$ and $n\ge 3$.

Then we choose 
\begin{equation}
p=\frac n2 -1
\end{equation}
and consider the equation
\begin{equation}\lae{2.11}
\begin{aligned}
&\frac n{16(n-1)}t^{-(m+k)}\frac\pa {\pa t}(t^{(m+k)}\dot u)\\
&\q -t^{-2}\D_Mu+\frac12 t^{-2}\D_{\R[k]}u-(n-1)t^{2-\frac4n}\tilde\D_\s u\\
&\q-(\frac n2-1)t^{2-\frac4n}R_\s u +(n-2)t^2\Lam u=0.
\end{aligned}
\end{equation}

\tit{Case 2}: $\Lam >0$ and $n\ge 5$.

Then, we choose
\begin{equation}
p=\frac n2 -2 -\frac14>0
\end{equation}
and consider the equation
\begin{equation}\lae{2.13}
\begin{aligned}
&-\frac14\frac n{16(n-1)}t^{-(m+k)}\frac\pa {\pa t}(t^{(m+k)}\dot u)\\
&\q +\frac14t^{-2}\D_Mu+\frac98 t^{-2}\D_{\R[k]}u-(n-1)t^{2-\frac4n}\tilde\D_\s u\\
&\q-(\frac n2-\frac94)t^{2-\frac4n}R_\s u +(n-\frac92)t^2\Lam u=0.
\end{aligned}
\end{equation}

\tit{Case 3}: $\Lam>0$ and $n=3$.

Then we choose
\begin{equation}
p=-\frac14
\end{equation}
yielding
\begin{equation}\lae{2.15}
\begin{aligned}
&\frac14\frac n{16(n-1)}t^{-(m+k)}\frac\pa {\pa t}(t^{(m+k)}\dot u)\\
&\q -\frac14t^{-2}\D_Mu+\frac78 t^{-2}\D_{\R[k]}u-(n-1)t^{2-\frac4n}\tilde\D_\s u\\
&\q+\frac14t^{2-\frac4n}R_\s u -\frac12t^2\Lam u=0.
\end{aligned}
\end{equation}
For a more detailed exposition we refer to \cite[Chapter 4.2]{cg:qgravity-book2}.

    Finally, let us look at the Wheeler-DeWitt equation which we solved when we quantized gravity combined with the forces of the Standard Model, \cf \cite{cg:uqtheory3b}. For our purpose the reference \cite[Chapter 5.4]{cg:qgravity-book2} is more suitable since, there, we also added a scalar field map such that the combined Hamilton function has the form
\begin{equation}\lae{2.9.4.81}
\begin{aligned}
\mc H&={\mc H}_G+ \mc H_S+{\mc H}_{YM}+{\mc H}_H+{\mc H}_D\\
&={\mc H}_G+\mc H_S+t^{-\frac23}(\tilde {\mc H}_{YM}+\tilde {\mc H}_H+\tilde {\mc H}_D)\\
&\equiv {\mc H}_G+\mc H_S+t^{-\frac23} \tilde {\mc H}_{SM},
\end{aligned}
\end{equation}
where  the subscripts $YM$, $H$, $D$ refer to the Yang-Mills, Higgs and Dirac fields and $SM$  to the fields  of the Standard Model or to a  corresponding subset of fields. The Hamilton constraint
\begin{equation}
\mc H=0
\end{equation}
will be quantized by first quantizing the Hamiltonians $\mc H_G+\mc H_S$ in the fibers for general metrics resulting in a hyperbolic operator
\begin{equation}
\hat{\mc H}_G+\hat{\mc H}_S
\end{equation}
But the expression
\begin{equation}
\hat{\mc H}_G u+\hat{\mc H}_S u
\end{equation}
will be evaluated $(x,t,\de_{ij}, \bar\theta^a)$, where $\de_{ij}$ is the standard Euclidean metric in $\so=\R[n]$, $n=3$, and
\begin{equation}
\bar\theta^a(x)=1\qq\A\, 1\le a\le k.
\end{equation}
The Hamilton function $\mc H_{SM}$, which represents spatial fields and is independent of $t$, is quantized in $(\so, \de_{ij})$ by the usual methods of  Quantum Field Theory (QFT). The Wheeler-DeWitt equation then has the form
\begin{equation}\lae{2.5.4.6.1.1}
\begin{aligned}
\hat{\mc H}u&=\al_N^{-1}\{\frac n{16(n-1)}t^{-(m+k)}\frac\pa {\pa t}(t^{(m+k)}\dot u)\\
&\qq -t^{-2}\D_Mu\}+\al_N^{-1}2t^2\Lam u+ t^{-\frac23}\hat{\mc H}_{SM}u=0,
\end{aligned}
\end{equation}
where $\al_N$ is a positive coupling constant and where we also assume that $u$ does not depend on $\theta^a(x)$.

We then solve the Wheeler-DeWitt equation by using separation of variables. The operator $\hat{\mc H}_{SM}$ is acting only in the base space $\so$, such that the spatial eigendistributions, or approximate eigendistributions, $\psi$ satisfying
\begin{equation}\lae{2.5.4.71.1}
\hat{\tilde {\mc H}}_{SM}\psi=\mu\psi,\qq \mu> 0
\end{equation}
can be derived by applying standard methods of QFT. 

 The remaining operator in \re{2.5.4.6.1.1}  is acting only in the fibers, i.e., we can use the eigenfunctions $v=v(\s_{ij})$ of $-\D_M$, which represent the elementary gravitons, satisfying 
\begin{equation}\lae{2.5.4.73.1}
-\D_Mv=(\abs\lam^2+\abs \rho^2)v\qq\A\,\s_{ij}\in M
\end{equation}
and\begin{equation}\lae{2.5.4.74.1}
v(\de_{ij})=1\qq\A\, x\in\so,
\end{equation}
\cf \cite[Theorem 3.2.3, p. 76]{cg:qgravity-book2}, and where
\begin{equation}\lae{2.5.4.75.1}
\abs\rho^2=1
\end{equation}
if $n=3$, compare \cite[equation (2.2.34), p. 49]{cg:qgravity-book2} and \fre{1.30}.

Hence, we make the ansatz 
\begin{equation}
u=wv\psi,
\end{equation}
where $w=w(t)$ only depends on $t>0$. Then, combining \re{2.5.4.6.1.1}, \re{2.5.4.71.1}, \re{2.5.4.73.1}, \re{2.5.4.74.1} and \re{2.5.4.75.1} we derive an ODE which  must be solved by $w$, namely,
\begin{equation}\lae{2.5.4.6.1.1.0}
\begin{aligned}
&\frac n{16(n-1)}t^{-(m+k)}\frac\pa {\pa t}(t^{(m+k)}\dot w)\\
&\qq +t^{-2}(\abs\lam^2+1)w+2t^2\Lam w+\al_Nt^{-\frac23}\mu w=0.
\end{aligned}
\end{equation}

Rewriting this ODE as
\begin{equation}\lae{2.28}
\begin{aligned}
-t^{-(m+k)}\frac\pa {\pa t}(t^{(m+k)}\dot w)
-\mu_0t^{-2}w-m_2t^2\Lam w=m_1t^{-\frac23} w,
\end{aligned}
\end{equation}
where
\begin{equation}
\begin{aligned}
\mu_0=\frac{16(n-1)}n(\abs\lam^2+1),
\end{aligned}
\end{equation}
\begin{equation}
m_1=\frac{16(n-1)}n\al_N\mu
\end{equation}
and
\begin{equation}
m_2=\frac{32(n-1)}n,
\end{equation}
then the left-hand side of \re{2.28} is identical to the left-hand side of equation \fre{1.31}. However, on the right-hand side of these equations we have different powers of $t$ which will lead to slightly different asymptotic estimates from above near the  origin for the corresponding solutions. In order to unify the approach we shall consider the temporal equation
\begin{equation}\lae{2.28.1}
\begin{aligned}
-t^{-(m+k)}\frac\pa {\pa t}(t^{(m+k)}\dot w)
-\mu_0t^{-2}w-m_2t^2\Lam w=m_1t^q w,
\end{aligned}
\end{equation}
where
\begin{equation}
-2<q<2
\end{equation}
such that the resulting estimates can be  applied in both cases.

Using the same transformation as in \fre{1.33} we define the function
\begin{equation}
u=t^{\frac{m+k-1}2} w
\end{equation}
which satisfies the equation
\begin{equation}\lae{2.35}
-t^{-1}\frac\pa{\pa t}\big (t \pde ut\big )-t^{-2}\bar\mu u
-t^2 m_2\Lam  u=m_1 t^q  u,
\end{equation}
where
\begin{equation}
\bar\mu=\mu_0-{\bigg(\frac{m+k-1}2\bigg)}^2
\end{equation}
 is negative if $k\in\N$ is large enough. If in addition the cosmological constant is also  negative
\begin{equation}
\Lam<0,
\end{equation} 
then \re{2.35} can be looked at as an eigenvalue equation with  positive eigenvalues $m_1$ in an appropriate Hilbert space. We shall solve the eigenvalue problem in the next section and prove asymptotic estimates near the singularity which will allow us to deduce that unitarily equivalent eigenfunctions can be extended past the singularity as sufficiently smooth functions.

\section{Extending the temporal solutions past the singularity}\las{3}

In this section we shall prove asymptotic estimates from above near the singularity for the solutions of the equation \re{2.35} and we shall use these estimates to conclude that the unitarily equivalent eigenfunction
\begin{equation}\lae{3.1}
\tilde u=t^\frac12 u
\end{equation}
can be extended past the singularity under suitable assumptions.

The extension itself is fairly easy we simply mirror the solution on the positive axes to the negative axes where even or odd mirroring are both possible. The crucial point is to show that the mirrored functions are sufficiently smooth in $\R[]$ and that the temporal equation is valid in the classical sense even at the singularity $t=0$. In order to achieve these results we have to prove that the temporal solutions and there derivatives, up to the order two at least, vanish sufficiently fast at $t=0$.

Next, let us prove sharp  estimates  near the origin for eigenfunctions of the equation \re{2.35} which will play a fundamental role in deducing that the unitarily equivalent temporal eigenfunctions $\tilde u$ in \re{3.1} which are the eigenfunctions of unitarily equivalent self-adjoint operator, can be smoothly extended past the big bang singularity in $t=0$. 

For a better understanding we first need a few definitions. The operator
\begin{equation}\lae{2.8.5.6}
B u=-t^{-1}\frac\pa{\pa t}\big (t \pde ut\big )+t^{-2}\mu^2 u
\end{equation}
is known as a Bessel operator.
\bd\lad{2.3.4.2.1}
Let $I=(0,\un)$  and let $r\in\R[]$. Then we define
\begin{equation}
L^2(I,r)=\set{u\in L^2_{\tup{loc}}(I,\R[])}{\int_It^r\abs u^2<\un}.
\end{equation}
$L^2(I,r)$ is a Hilbert space with scalar product 
\begin{equation}
\spd{u_1}{u_2}_r=\int_I t^ru_1u_2,
\end{equation}
but let us emphasize that we shall apply this definition only for $r\not=2$. The scalar product
$\spd\cdot\cdot_2$ will be defined differently.
\ed
We consider real valued functions for simplicity but we could just as well allow complex valued functions with the standard scalar product, or more precisely, sesquilinear form.
\bd\lad{2.8.5.2.1}
For functions $u\in C^\un_c(I)$ define the operator
\begin{equation}
A_1u=-t^{-1}\frac\pa{\pa t}\big (t \pde ut\big )+t^{-2}\mu^2 u
- t^2 m_2 \Lam u,
\end{equation}
as well as the scalar product 
\begin{equation}\lae{3.6}
\spd{u_1}{u_2}_2=\spd{Bu_1+t^2m_2u_1}{u_2}_1\qq\A\, u_1,u_2\in C^\un_c(I).
\end{equation}
\ed
The right-hand side of \re{3.6} is an integral. Integrating by parts we deduce
\begin{equation}\lae{2.8.5.12.1.1}
\spd{u_1}{u_2}_2=\int_I(t\dot u_1\dot u_2+\mu^2 t^{-1}u_1u_2+t^3m_2 u_1u_2),
\end{equation} 
i.e., the scalar product is indeed positive definite because of the assumption $\mu>0$. Let us define the norm
\begin{equation}\lae{2.8.5.13} 
\norm u_2^2=\spd uu_2\qq \A\, u\in C^\un_c(I)
\end{equation}
and the Hilbert space $\mc H_2=\mc H_2(I)$ as the closure of $C^\un_c(I)$ with respect to the norm $\norm\cdot_2$.
\bpp
The functions $u\in \mc H_2$ have the properties
\begin{equation}\lae{2.8.5.15.1}
u\in C^0([0,\un)),
\end{equation}
\begin{equation}\lae{2.8.5.16.1}
\abs {u(t)}\le c\norm u_2\qq\A\,t\in I,
\end{equation}
where $c=c(\mu, m_2,\abs\Lam)$,
\begin{equation}\lae{2.8.5.17.1}
\lim_{t\ra 0} u(t)=0
\end{equation} 
and
\begin{equation}\lae{2.8.5.18.1}
\abs{u(t)}\le c \norm u_2 t^{-1}\qq\A\, t\in I,
\end{equation}
where $c$ is a different constant depending on $\mu, m_2$ and $\abs\Lam$.
\epp
For a proof we refer to \cite[Proposition 3.4.3, p. 82]{cg:qgravity-book2}.

\bt\lat{2.3.4.10.1}
Let  $u\in \mc H_2$ satisfy the equation
\begin{equation}\lae{2.3.4.47.1}
A_1u=-t^{-1}\frac\pa{\pa t}\big (t \pde ut\big )+t^{-2}\mu^2 u
+t^2 m_2^2  u=\lam t^q  u,
\end{equation}
where the constants $\mu, m_2$ and $\lam$ are strictly positive and the exponent $q$ satisfies 
\begin{equation}\lae{3.2}
-2<q<2.
\end{equation}
Since $\mu$ is especially important, let us emphasize that  
\begin{equation}\lae{3.3}
\mu^2=-\bar\mu=\frac{(m+k-1)^2}4+\ga_0 \abs{\theta_0}^2-\mu_0,
\end{equation}
where $\ga_0$ is a positive constant, $\theta_0\in \R[k]$ an arbitrary, but fixed, vector
and $\mu_0>0$. Then, for every $\e>0$  there exists $0<t_0<1$ and a positive constant $c_1$ such that $u$ does not vanish in the interval $(0,t_0]$ and can be estimates by
\begin{equation}\lae{3.16}
\abs {u(t)}\le c_1 t^{\mu_\e}\qq\A\,t\in (0,t_0],
\end{equation}
where $0<\mu_\e$ is defined by 
\begin{equation}\lae{3.17}
\mu_\e^2=\mu^2-\e>0.
\end{equation}
\et
\bp
Let us first prove that $u$ does not vanish for small $t>0$. Arguing by contradiction let $0<t_0<1$ be a point where
\begin{equation}
u(t_0)=0.
\end{equation}
Multiplying the equation \re{2.3.4.47.1} by $t u$ and integrating by parts over the interval $[0,t_0]$ we infer
\begin{equation}
\int_0^{t_0}\mu^2 t^{-1}\abs u^2\le \int_0^{t_0}\lam t^{1+q}\abs u^2
\end{equation}
and conclude further that $t_0$ cannot be arbitrarily close to $0$.

Thus, let us assume $u$ to be real valued and strictly positive in $(0,t_0]$ for some small $t_0$. To prove the inequality in \re{3.16}, let us consider the equation
\begin{equation}\lae{2.3.4.60.1}
A_{1,\e} \psi=\lam \psi\qq\tup{in}\; (0,\un)
\end{equation}
requiring
\begin{equation}
\psi(0)=0,
\end{equation}
 where the operator $A_{1,\e}$ is defined by replacing $\mu$ by $\mu_\e$ in  equation \re{2.3.4.47.1}.
One can easily verify that a solution $\psi=\psi(t)$ satisfying both equations is given by defining
\begin{equation}\lae{2.3.4.62.1}
\psi(t)=e^{-\frac12 m_2t^2}t^{\mu_\e} M(a,b,m_2t^2),
\end{equation}
where
\begin{equation}\lae{3.23.1}
a=\frac12 (\mu_\e+1)-\frac14\frac \lam{m_2}
\end{equation}
and
\begin{equation}
b=\mu_\e+1.
\end{equation}
$M=M(a,b,z)$, $z\in\Cc$,  is known as \tit{Kummers's function} or as the entire \tit{confluent hypergeometric function}  which is a solution of \tit{Kummer's equation}
\begin{equation}
zy''+(b-z)y'-ay=0
\end{equation}
and which can be expressed by the power series  
\begin{equation}
_1F_1(a,b,z)=M(a,b,z)=1+\frac ab z+\sum_{k=2}^\un \frac{a(a+1)\cdots (a+k-1)z^k}{b(b+1)\cdots (b+k-1)k!}
\end{equation}
which is absolutely convergent for any $z\in\Cc$ provided
\begin{equation}
b\not\in \Z_{\le 0},
\end{equation}
which is certainly true in our case. For a detailed analysis of the solutions of Kummer's equation we refer to \cite[Chapter 13.2, p. 322]{nist:handbook} or \cite[p. 427]{kamke:ode}. 

Obviously $u$ is a subsolution of the equation \re{2.3.4.60.1} in the interval $(0,t_0)$, i.e.,  
\begin{equation}
A_{1,\e}u\le \lam u,
\end{equation}
because
\begin{equation}
A_{1,\e}u=\lam t^q u-\e t^{-2}u<0
\end{equation}
if $t_0$ is small enough.
Moreover, $\psi(t)$ is positive if $t_0$ is small, for $M(a,b,0)=1$,  hence, there exists a constant $c_2$ such that
\begin{equation}
u(t_0)=c_2 \psi(t_0).
\end{equation}
In order to prove \re{3.16} we multiply the inequality
\begin{equation}
A_{1,\e}(u-c_2\psi)\le \lam (u-c_2\psi)
\end{equation}
 by $t\max(u-c_2\psi,0)$ and  partially integrating the result  in the interval $(0,t_0]$ yields
\begin{equation}
\int_0^{t_0}\mu_\e^2t^{-1}\max(u-c_2\psi,0)^2\le \int_0^{t_0}\lam \max(u-c_2\psi,0)^2
\end{equation}
from which we deduce
\begin{equation}
u(t)\le c_2 \psi(t)\qq\A\, t\in [0,t_0]
\end{equation}
if $t_0$ is small, completing the proof of the theorem,  in view of  the definition of $\psi$ in \re{2.3.4.62.1}. 
\ep 

\br
The assumptions regarding the coefficients and the exponents of the ODE in the theorem above cover the cases we are confronted with after the quantization of the full Einstein equations, where $\theta_0\in\R[k]$ and $n\ge 3$ can be arbitrary and $q=2-\frac2n$, as well as in case of the Wheeler-DeWitt equation, where we have to choose $\theta_0=0$, n=3 and $q=-\frac23$. To ensure that the right-hand side of equation \re{3.3} is positive in the latter case the dimension $k$ of the target space of the scalar field map, which is $\R[k]$,  has to be sufficiently large.   
\er

We shall apply the estimate \re{3.16} to the function
\begin{equation}\lae{3.22}
\tilde u=t^\frac12 u,
\end{equation}
which satisfies the differential equation
\begin{equation}\lae{3.23}
-\Ddot{ \tilde u}  + t^{-2} {\tilde \mu}^2 \tilde u +t^2 m_2^2  \tilde u=\lam t^q  \tilde u,
\end{equation}
where 
\begin{equation}\lae{3.36}
\tilde\mu^2=\mu^2-\frac14,
\end{equation}
as can be easily checked. 

But before we shall prove that the eigenfunctions in equation \re{3.23} can be extended past the singularity as sufficiently smooth functions, let us verify that equation \re{3.23} is unitarily equivalent to equation \re{2.3.4.47.1}, if we consider complex valued functions, otherwise there is an orthogonal equivalence. After that verification the countably many eigenfunctions $\tilde u_i$ with eigenvalues $\lam_i$ can be looked at as the temporal eigenfunctions of our model of quantum gravity which can be extended past the singularity. We shall also  prove that equation \re{3.23} can be defined as a classical equation for $(\tilde u_i, \lam_i)$ valid in $\R[]$ provided $t^q$ is replaced by $\abs t^q$ and $\tilde u_i$ is extended by reflection either even or odd.

\bd\lad{2.8.5.2}
For functions $u\in C^\un_c(I)$ define the operators
\begin{equation}
 A_ru=-t^{-r}\frac\pa{\pa t}\big (t^r \pde u t\big )+t^{-2}\mu^2u
+ t^2 m_2  u
\end{equation}
and  the scalar product 
\begin{equation}\lae{3.38.1}
\spd{u_1}{u_2}_2=\spd{A_ru_1}{u_2}_r\qq\A\, u_1,u_2\in C^\un_c(I). 
\end{equation}
\ed
The right-hand side of \re{3.38.1} is an integral. Integrating by parts we deduce 
\begin{equation}\lae{2.8.5.12.1}
\spd{u_1}{u_2}_2=\int_I(t^r\dot u_1\dot u_2+\mu^2 t^{r-2}u_1u_2+t^{r+2}m_2 u_1u_2).
\end{equation} 
 Let us define the norm
\begin{equation}\lae{2.8.5.13.1}
\norm u_2^2=\spd uu_2\qq \A\, u\in C^\un_c(I)
\end{equation}
and the Hilbert space $\mc H_2=\mc H_2(I)$ as the closure of $C^\un_c(I)$ with respect to the norm $\norm\cdot_2$.

Define the operator $A_0$ in $C^\un_c(I)$ by 
\begin{equation}
A_0=-\Ddot{\tilde u}+t^{-2} \tilde\mu^2\tilde u+t^2m_2 \tilde u,
\end{equation}
where
\begin{equation}
\tilde\mu^2 =\mu^2+\frac{r^2}4-\frac r2
\end{equation}
is supposed to be strictly positive and let $\tilde {\mc H}_2$ be the completion with respect to the corresponding norm
\begin{equation}\lae{3.33}
\nnorm {\tilde u}^2=\int_0^\un (\abs{\dot{\tilde u}}^2+t^{-2}\tilde\mu^2\tilde u^2+t^2m_2\tilde u^2)=\spd{A_0\tilde u}{\tilde u}\equiv \spdd{\tilde u}{\tilde u}_2
\end{equation}
\bpp
The functions $\tilde u\in \mc {\tilde H}_2$ have the properties
\begin{equation}\lae{2.8.5.15}
\tilde u\in C^0([0,\un)),
\end{equation}
\begin{equation}\lae{2.8.5.16}
\abs {\tilde u(t)}\le c\nnorm{\tilde u}_2\qq\A\,t\in I,
\end{equation}
where $c=c(\bar\mu, m_2)$,
\begin{equation}\lae{2.8.5.17}
\abs{\tilde u(t)}\le c \nnorm {\tilde u}_2 t^{\frac12}\qq\A\, t\in I,
\end{equation}
as well as
\begin{equation}\lae{2.8.5.18}
\abs{\tilde u(t)}\le c \nnorm {\tilde u}_2 t^{-\frac12}\qq\A\, t\in I,
\end{equation}
where $c$ is a different constant depending on $\tilde\mu, m_2$. 
\epp
\bp
Let us first assume $\tilde u\in C^\un_c(I)$ and let $\de>0$, then
\begin{equation}\lae{3.38}
\tilde u^2(\de)=2\int_0^\de\dot {\tilde u }\tilde u\le\int_0^\de \abs{\dot {\tilde u}}^2+\int_0^\de \abs {\tilde u}^2.
\end{equation}
This estimate is  also valid for any $\tilde u\in \mc {\tilde H}_2$ by approximation which in turn implies the relations \re{2.8.5.16}.

Next let us slightly modify the  previous argument to obtain
\begin{equation}\lae{3.39}
\tilde u^2(\de)=2\int_0^\de\dot {\tilde u }\tilde u\le2\bigg(\int_0^\de \dot{\tilde u}^2\bigg)^\frac12 \bigg(\int_0^\de t^{-2}t^2{\tilde u}^2\bigg)^\frac12\le c \nnorm {\tilde u}_2^2 \,\de
\end{equation}
from which we infer \re{2.8.5.17} and also \re{2.8.5.15} since $u$ is  continuous in $I$. 

It remains to prove \re{2.8.5.18}. Let $\tilde u\in \mc{\tilde H}_2$ and define $\tilde v=\tilde v(\tau)$ by
\begin{equation}
\tilde v(\tau)=\tilde u(\tau^{-1}),
\end{equation}
where $\tau=t^{-1}$ for all $t>0$. Applying simple calculus arguments we then obtain
\begin{equation}\lae{2.8.5.21}
\begin{aligned}
\int_0^\un\{\tau^2\abs{\tilde v'}^2+\tau^2\tilde\mu^2\abs{\tilde v}^2+\tau^{-4}m_2 \abs {\tilde v}^2\}d\tau=\nnorm {\tilde u}_2^2
\end{aligned}
\end{equation}
as well as  
\begin{equation}\lae{2.8.5.22}
\begin{aligned}
\int_0^\un\{\tau^2\abs{\tilde v'}^2+\tau^2{\tilde\mu}^2\abs {\tilde v}^2\}d\tau=\int_0^\un\{\abs{\dot{\tilde u}}^2+t^{-2}{\tilde\mu}^2 \abs {\tilde u}^2\}dt.
\end{aligned}
\end{equation}
Moreover, first assuming, as before, that $\tilde u$ and hence $\tilde v$ are test functions we argue as in \re{3.39} that for any $\de>0$
\begin{equation}
\begin{aligned}
\tilde v^2(\de)=2\int_0^\de \tilde v'\tilde v&\le 2\bigg(\int_0^\de \tau^2 \abs{\tilde v'}^2\bigg)^\frac12\bigg(\int_0^\de \tau^{-2} \abs{\tilde v}^2\bigg)^\frac12\\
&\le  2\bigg(\int_0^\de \tau^2 \abs{\tilde v'}^2\bigg)^\frac12\bigg(\int_0^\de \tau^{-4} \abs{\tilde v}^2\bigg)^\frac12 \de\\
&\le c \nnorm{ \tilde u}_2^2\,\de,
\end{aligned}
\end{equation}
where we used \re{2.8.5.21} for the last inequality and where $c=c(\tilde\mu,m_2)$. Setting $\de=t^{-1}$ for arbitrary $t>0$ we have proved the estimate \re{2.8.5.18} for test functions and hence for arbitrary $\tilde u\in \tilde{\mc H}_2$.  
\ep

\bl\lal{3.6}
Assuming the definitions in \rd{2.8.5.2}, then the map
\begin{equation}
\begin{aligned}
\f:\mc H_2&\ra \tilde{\mc H}_2,\\
u&\ra \tilde u=t^\frac r2 u
\end{aligned}
\end{equation}
is orthogonal if the functions are supposed to be real valued and unitary if complex functions are considered and the scalar products are suitably modified, i.e,
\begin{equation}\lae{3.45}
\spd {A_r u_1}{u_2}_r=\spd{A_0\tilde u_1}{\tilde u_2}\qq\A\, u_i\in \mc H_2, \;i=1,2,
\end{equation}
and 
\begin{equation}\lae{3.46}
\begin{aligned}
A_r&=\f^{-1}\circ A_0\circ\f,\\
A_0&=\f\circ A_r\circ \f^{-1},
\end{aligned}
\end{equation}
i.e., $A_r$ and $A_0$ are unitarily equivalent.
\el
\bp
For the prove of \re{3.45} we may assume that the functions are real valued. The relation is then easily verified by applying elementary calculus:
\begin{equation}
\begin{aligned}
t^r \dot u_1\dot u_2=\dot{\tilde u}_1\dot{\tilde u}_2+ \frac {r^2}4t^{-2} \tilde u_1\tilde u_2-\frac r2t^{-1}(\tilde u_1\tilde u_2)'
\end{aligned}
\end{equation}
from which we deduce by applying partial integration
\begin{equation}
\begin{aligned}
&\int_0^\un \{t^r \dot u_1\dot u_2+\mu^2t^{r-2}u_1u_2+m_2 t^{r+2}u_1u_2\}=\\
&\q \int_0^\un\{\dot{\tilde u}_1\dot{\tilde u}_2+(\mu^2+\frac{r^2}4-\frac r2)t^{-2}\tilde u_1\tilde u_2+m_2t^2\tilde u_1\tilde u_2\}.
\end{aligned}
\end{equation}

Moreover, a straightforward calculation reveals that for test functions $u$
\begin{equation}
A_0\tilde u=t^\frac r2 A_r u
\end{equation}
proving \re{3.46}.
\ep
The equations \re{3.23} \resp
\begin{equation}\lae{3.50}
A_r=\lam t^q  u
\end{equation}
can be looked at as eigenvalue equations which can be expressed abstractly in the form: $u\in \mc H_2$ satisfies
\begin{equation}\lae{3.51}
B(u,v)\equiv \spd{A_r u}v_r=\lam K(u,v)\qq\A\, v\in \mc H_2,
\end{equation}
where
\begin{equation}
K(u,v)=\int_0^\un t^{r+q} uv
\end{equation}
and where we only consider real valued functions for simplicity. Since $r\in \R[]$ is arbitrary the case $r=0$ is also covered. 

\bt
The eigenvalue problem \re{3.51} is orthogonally (unitarily) equivalent to the corresponding eigenvalue problem: $\tilde u\in \tilde{\mc H}_2$ satisfies
\begin{equation}\lae{3.53}
\tilde B(\tilde u,\tilde v)\equiv \spd{A_0\tilde u}{\tilde v}=\lam \tilde K(\tilde u,\tilde v) \qq \A\, \tilde v\in \tilde{\mc H}_2,
\end{equation}
where
\begin{equation}\lae{3.54} 
\tilde K(\tilde u,\tilde v)=\int_0^\un t^q \tilde u \tilde v.
\end{equation} 
Hence, the respective eigenvalues are identical.
\et
\bp
Let $\f$ be the unitary map in \rl{3.6}. In view of \re{3.45} we conclude
\begin{equation}
B(u,v)=\tilde B(\f(u),\f(v))
\end{equation}
and also
\begin{equation}
K(u,v)=\tilde K(\f(u),\f(v))
\end{equation}
completing the proof.
\ep

If $q\in \R[]$ satisfies the estimates 
\begin{equation}\lae{3.57}
-2< q<2
\end{equation}
then the quadratic form
\begin{equation}
\tilde K(v)=\tilde K(v,v) 
\end{equation}
is \tit{compact} with respect to the quadratic form
\begin{equation}
\tilde B(\tilde v)=\tilde B(\tilde v,\tilde v),
\end{equation}
 i.e., if
\begin{equation}
\tilde B(\tilde v_i,\tilde v)\ra \tilde B(\tilde v_0),\tilde v)\qq\A\, \tilde v\in \tilde {\mc H}_2
\end{equation}
then
\begin{equation}
\tilde K(\tilde v_i-\tilde v_0)\ra 0.
\end{equation}
The proof is well-known and fairly simple: In compact subintervals of $(0,\un)$ the compactness follows from the Sobolev embedding theorems and near the endpoints of the interval $t=0$ and $t=\un$ the compactness can be deduced from the finiteness of
\begin{equation}
\int_0^\un (t^{-2}+t^2)\abs{\tilde v_i -\tilde v_0}^2 \le \const\qq\A\, i\in\N.
\end{equation}
The latter estimate is due to the definition of the scalar product $\tilde B$ and the uniform boundedness principle which says that any weakly bounded sequence in a Banach space is uniformly bounded.

If these conditions are satisfied then the following theorem is well-known: 
\bt\lat{3.8}
The eigenvalue problem \re{3.53} has countably many solutions $(\lam_i,\tilde u_i)$, $\tilde u_i\in \tilde{\mc H}_2$, with the properties
\begin{equation}
\lam_i<\lam_{i+1}\qq\A\,i\in\N,
\end{equation}
\begin{equation}
\lim_i\lam_i=\un,
\end{equation}
\begin{equation}
\tilde K(\tilde u_i,\tilde u_j)=\de_{ij}.
\end{equation}
 The pairs $(\lam_i, \tilde u_i)$ are  recursively defined  by the variational problems
 \begin{equation}\lae{3.66}
\lam_0=\tilde B(\tilde u_0)=\inf \bigg\{\frac{\tilde B(u)}{\tilde K(u)}:0\not=u\in \tilde{\mc H}_2\bigg\}
\end{equation}
and for $i>0$
 \begin{equation}
\lam_i=\tilde B(\tilde u_i)=\inf \bigg\{\frac{\tilde B(u)}{\tilde K(u)}:0\not=u\in \tilde{\mc H}_2,\, \tilde K(u,u_j)=0, \, 0\le j\le i-1\bigg\}.
\end{equation}
 The $(\tilde u_i)$ form a Hilbert space basis in $\tilde{\mc H}_2$ and in $L^2(I,q)$, the eigenvalues are strictly  positive and the eigenspaces are one dimensional. 
\et
\bp
This theorem is well-known and goes back to the book of Courant-Hilbert \cite{courant-hilbert-I}, though in a general separable Hilbert space the eigenvalues are not all positive and the eigenspaces are only finite dimensional. For a proof in the general case we refer to \cite[Theorem 1.6.3, p. 37]{cg:pdeII}. 

The positivity of the eigenvalues in the above theorem is obvious and the fact that the eigenspaces are one dimensional is proved by contradiction. Thus, suppose there exist an eigenvalue $\lam=\lam_i$ and two corresponding linearly independent eigenfunctions $\tilde u_1,\tilde u_2\in \tilde{\mc H}_2$. Then, for any $t_0>0$ there would exist an eigenfunction $u\in \tilde{\mc H}_2$ with eigenvalue $\lam$  satisfying $u(t_0)=0$ and the equation \re{3.53}. Multiplying this equation by $u$ and integrating the result in the interval $(0,t_0)$ with respect to the measure $dt$ we obtain
\begin{equation}
\int_0^{t_0}\tilde\mu^2 t^{-2} u^2\le t_0^{2+q}\int_0^{t_0}\lam  t^{-2} u^2,
\end{equation}
where we used
\begin{equation}
1\le \frac{t_0}t,\qq \A\,t\in (0,t_0),
\end{equation}
and
\begin{equation}
2+q>0,
\end{equation}
in view of \re{3.2}, yielding a contradiction if $t_0$ is sufficiently small.
\ep

\br
The previous results are a also valid if instead of the coefficient
\begin{equation}
m_2 t^2
\end{equation}
we consider the actual coefficient
\begin{equation}
m_2 \abs\Lam t^2,
\end{equation}
where in our case $\Lam<0$. The eigenvalues $\lam_i$ then depend on $\Lam$.
\er
In \cite[Lemma 9.4.8, p. 240]{cg:qgravity-book2} we proved the following lemma, which we include here together with an appropriately modified proof for the convenience of the reader.
\bl\lal{3.10}
Let $\lam_i$ be the temporal eigenvalues depending on $\Lam$ and let $\bar\lam_i$ be the corresponding eigenvalues for
\begin{equation}
\abs\Lam=1,
\end{equation}
then
\begin{equation}\lae{3.74}
\lam_i=\bar\lam_i\abs\Lam^\frac{2+q}4.
\end{equation}
\el
\bp
Let $\tilde B$ and $\tilde K$ be the quadratic forms defined by
\begin{equation}\lae{3.75}
\tilde B(u)=\int_0^\un\{\abs{\dot u}^2+t^{-2}\abs{\tilde\mu}^2\abs u^2+t^2m_2\abs\Lam \abs u^2\}
\end{equation}
and
\begin{equation}
\tilde K(u)=m_3\int_0^\un t^{q}\abs u^2 
\end{equation}
and let $\tilde B_1(u)$ the quadratic form by choosing $\abs\Lam=1$ in $\tilde B$.
 Then we have
\begin{equation}\lae{3.77}
\frac{\tilde B(u)}{\tilde K(u)}=\abs\Lam^\frac{2+q}4\frac{\tilde B_1(u)}{\tilde K(u)}\qq\A\,0\not=u\in C^\un_c(\R[]_+).
\end{equation}
To prove \re{3.77} we introduce a new integration variable $\tau$ on the left-hand side
\begin{equation}
t=\mu\tau,\qq\mu>0,
\end{equation}
to conclude
\begin{equation}
\frac{\tilde B(u)}{\tilde K(u)}=\mu^{-(2+q)}\frac{\tilde B_1(u)}{\tilde K(u)}\qq\A\,0\not=u\in C^\un_c(\R[]_+).
\end{equation}
provided
\begin{equation}
\mu=\abs\Lam^{-\frac14}.
\end{equation}
The relation \re{3.77} immediately implies \re{3.74}, in view of \rt{3.8}. 
\ep
\br\lar{3.11}
Let $(\tilde u_i,\lam_i)$ be the previous eigenfunctions and eigenvalues of the operator
\begin{equation}
-\Ddot{ \tilde u}  + t^{-2} {\tilde \mu}^2 \tilde u +t^2 m_2^2 \abs\Lam \tilde u
\end{equation}
with respect to the quadratic form
\begin{equation}
\tilde K(\tilde u)=\int_0^\un  t^q \abs{\tilde u}^2,
\end{equation}
define
\begin{equation}
\f_0(t)= t^q
\end{equation}
and let $H_0$ be the operator  
\begin{equation}\lae{3.84}
\f_0^{-1}(-\Ddot {\tilde u}  + t^{-2} {\tilde \mu}^2 \tilde u +t^2 m_2^2 \abs\Lam \tilde u)
\end{equation}
defined in the dense subspace of the Hilbert space $\mc H=L^2(I,\f_0 dt)$ generated by the eigenfunctions $(\tilde u_i)$, then $H_0$ is essentially self-adjoint and its closure, which we denote by the same symbol, is self-adjoint; for a proof see the remarks following \cite[Definition 3.4.14, p.91]{cg:qgravity-book2}.  
\er

In the next section we shall prove that for any $\bet>0$ 
\begin{equation}
e^{-\bet H_0}
\end{equation}
is of trace class in $\mc H$, i.e.,
\begin{equation}
\tr(e^{-\bet H_0})=\sum_{i=0}^\un e^{-\bet \lam_i}<\un.
\end{equation}
Because we consider arbitrary $q$ satisfying
\begin{equation}\lae{3.87}
-2<q<2
\end{equation}
and not only the special values
\begin{equation}
q=2-\frac2n\q \vee\q q=-\frac23
\end{equation}
we cannot refer to a previous result and an extra proof is necessary. 

After having established that $\tilde u$ is unitarily equivalent to the solution $u$ of \re{2.3.4.47.1} which in turn is unitarily equivalent to the solution $w$ of equation \fre{1.31} \resp \fre{2.28}, \cf \cite[Lemma 3.4.10, p. 89]{cg:qgravity-book2}, we shall consider the equation \re{3.23} and its solution $\tilde u$, defined in \re{3.22}, to be the temporal eigenfunction equation which we shall extend past the singularity. In view of the estimate \re{3.16}, where $\mu_\e$ is defined in \re{3.17} we infer, by using the fact that we may assume $u$ to be positive in $(0,t_0]$,  
\begin{equation}\lae{3.99}
0<\tilde u \le c_1 t^{\mu_\e+\frac12}\qq\A\, t \in (0,t_0],
\end{equation}
where $\e>0$ is as small as we like but fixed. The constant $c_1$ depends on $\e$ and will tend to infinity if $\e$ tends to zero. However, we are able to conclude
\bl\lal{3.12}
Let $1\le m_0\in \N$ and assume
\begin{equation}\lae{3.90}
\mu +\frac12 >m_0,
\end{equation}
then there exists $\e>0$ and positive constants $c_1, t_0$ such  that
\begin{equation}\lae{3.91}
0<\tilde u \le c_1 t^{m_0+\e}\qq\A\, t \in I=(0,t_0].
\end{equation}
\el
The proof is obvious.
\bl\lal{3.13}
Let  the assumption \re{3.90} be satisfied for $m_0=1$, then 
\begin{equation}\lae{3.92}
\tilde u \in C^1[0,t_0]\q\wed\q \dot{\tilde u}(0)=0.
\end{equation}
Moreover, $\tilde u$ is strictly convex in $[0,t_0]$ if $t_0$ is small enough. Extending $\tilde u$ to $[-t_0,0)$ by defining   
\begin{equation}
\tilde u(t)=
\begin{cases}
\tilde u(t),& t\ge 0,\\
\tilde u(-t), &t<0,
\end{cases}
\end{equation}
then the extended function is of class $C^1$ in $[-t_0,t_0]$, strictly convex and
\begin{equation}\lae{3.94}
\Ddot{\tilde u}>0
\end{equation}
in the distributional sense, i.e.,
\begin{equation}\lae{3.95}
\spd{\tilde u}{\Ddot\h}>0\qq\A\, 0\le \h\in C^\un_c(-t_0,t_0),
\end{equation}
which do not vanish identically.
\el
\bp
From the equation \re{3.23} we deduce 
\begin{equation}
\Ddot{\tilde u}(t)>0\qq\A\, t\in (0,t_0),
\end{equation}
if $t_0$ is small enough, hence $\tilde u$ is strictly convex in the interval. Since $\tilde u> 0$ and and $\tilde u(0)=0$, we infer
\begin{equation}
\dot{\tilde u}(t)>0\qq\A\, t\in I,
\end{equation}
because $\dot{\tilde u}$ is also monotone increasing. Hence, we conclude
\begin{equation}
0\le c=\lim_{t\ra 0}\dot{\tilde u}(t)
\end{equation}
exists. If $c>0$ we would obtain a contradiction in view of  \re{3.91}, i.e., the right derivative of $\tilde u$ satisfies
\begin{equation}
\dot{\tilde u}(0)=\lim_{t\ra 0}\frac{\tilde u(t)}t=0=\lim_{t\ra 0}\dot{\tilde u}(t),
\end{equation} 
hence, we  have proved \re{3.92}.

Finally, \re{3.94} is valid for any $0\not=t\in(-t_0,t_0)$ and the relation \re{3.95} follows by partial integration over the open subintervals $\{t\not=0\}$ by using
\begin{equation}
\tilde u(0)=0=\dot{\tilde u}(0).
\end{equation}
\ep

\bl
Let  the assumption \re{3.90} be satisfied for $m_0=2$, then  
\begin{equation}
\tilde u\in C^2([0,t_0]),
\end{equation}
\begin{equation}\lae{3.102}
\lim_{t\ra 0}\frac{{\tilde u}(t)}{t^2}=0
\end{equation}
and
\begin{equation}\lae{3.103}
\Ddot{\tilde u}(0)=0=\lim_{t\ra 0}\Ddot{\tilde u}(t)\q\wed\q \lim_{t\ra 0}\frac{{\dot{\tilde u}}(t)}t=0.
\end{equation}
Moreover, these properties are also valid for the extended function. 
\el
\bp
The equation \re{3.102} is valid due to \re{3.91}, while the first relation in \re{3.103} immediate follows from \re{3.102} and the equation satisfied by $\tilde u$.

To prove the second equation in \re{3.103} we apply De L'Hospital's rule and use the first equation. Finally, it is obvious that these properties are also valid for the extended function.
\ep
 We are now able to prove by induction
 \bt\lat{3.15}
 Let  the assumption \re{3.90} be satisfied for arbitrary $2\le m_0\in\N$, then  
 \begin{equation}\lae{3.104}
\tilde u\in C^{m_0}([0,t_0])\q\wed\q \tilde u^{(m_0)}(0)=0=\lim_{t\ra 0}\tilde u^{(m_0)}(t)
\end{equation}
as well as
\begin{equation}\lae{3.105}
\lim_{t\ra 0}\frac{\tilde u^{(k)}(t)}{t^{m_0-k}}=0\qq\A\, 1\le k\le m_0, \,k\in \N,
\end{equation}
where $\tilde u^{(k)}$ denotes the $k$-th derivative of $\tilde u$. These properties are also valid for the extended function. 
 \et
\bp
The claims in \re{3.104} are certainly correct provided the relations in \re{3.105} are valid. Hence, it suffices
to prove the relations in  \re{3.105} per induction with respect to $k$. Let us first consider the case $k=1$. Applying De L'Hospital's rule we deduce
\begin{equation}
\lim_{t\ra 0}\frac{\dot{\tilde u}(t)}{t^{m_0-1}}=(m_0-1)^{-1}\lim_{t\ra 0}\frac{\Ddot{\tilde u}(t)}{t^{m_0-2}}=(m_0-1)^{-1} \lim_{t\ra 0}\frac{\tilde u(t)}{t^{m_0}}=0,
\end{equation}
where we used for the second equality the equation satisfied by $\tilde u$ and for the last the estimate \re{3.91}. The last two arguments also reveal that the claim in \re{3.105} is true for $k=2$.

Thus, let us assume that the limit relations in \re{3.105} are  already valid for $1\le k\le p<m_0$ , $p\ge 2$, and let us prove that then they are also satisfied for $k=p+1$. Let us recall that $\tilde u$ is a solution of the equation \re{3.23} which we can write in the form
\begin{equation}
\Ddot{\tilde u}=\tilde \mu ^2 t^{-2}\tilde u+ (m_2^2 t^2-\lam t^q)\tilde u.
\end{equation} 
Differentiating both sides with respect to $D^{p-1}$, where $D$ denotes differentiation with respect to $t$, we deduce, by applying the product rule, 
\begin{equation}
\tilde u^{(p+1)}=\tilde\mu^2\sum_{k=0}^{p-1} c_{p,k} t^{-2-k}\tilde u ^{(p-1-k)}+R_1+R_2,
\end{equation}
where the additional terms $R_1, R_2$ have a similar structure as the detailed sum, but the exponents of $t$ are less critical for small $t>0$ than in the first sum. The arguments we shall use in the case of the first sum will also apply in case of the additional terms and will therefore be omitted.

Next, we have to prove
\begin{equation}\lae{3.109}
\lim_{t\ra 0}\frac{\tilde u^{(p+1)}}{t^{m_0-(p+1)}}=0.
\end{equation}
Indeed, we infer
\begin{equation}
\sum_{k=0}^{p-1} c_{p,k} \frac{\tilde u ^{(p-1-k)}}{t^{2+k+m_0-(p+1)}}=\sum_{k=0}^{p-1} c_{p,k} \frac{\tilde u ^{(p-1-k)}}{t^{m_0-(p-1-k)}}
\end{equation}
and the right-hand side converges to zero if $t$ tends to zero, in view of the induction assumption. Hence, the relation \re{3.109} is proved completing the proof of the theorem.
\ep
As a corollary we obtain
\bc\lac{3.18}
If the assumption of the preceding theorem is satisfied then the extended solution $\tilde u$ also satisfies the extended equation 
\begin{equation}\lae{3.111}
-\Ddot{ \tilde u}  + t^{-2} {\tilde \mu}^2 \tilde u +t^2 m_2^2  \tilde u=\lam \abs t^q  \tilde u
\end{equation}
in $\R[]$, where we have to replace $t^q$ by $\abs t^q$ for obvious reasons. Let us emphasize that the lower order coefficients of the ODE exhibit a singularity in $t=0$ but that both sides of the equation are continuous in the interval $(-\un,\un)$ and vanish in $t=0$.
\ec
Let us conclude this section with a qualitative plot of an eigenfunction of equation \re{2.3.4.47.1} and with a sharp estimate near infinity for a solution of the equation \re{3.111}. For the plot of an eigenfunction we unfortunately can not consider one of the exponents
\begin{equation}
q=2-\frac4n\q \vee  \q q=-\frac23,   
\end{equation}
we are interested in, but only the exponent $q=0$ because then an eigenfunction is given by the function $\psi$ in \re{2.3.4.62.1} which has known eigenvalues. These are characterized by requiring that Kummer's function can be expressed as a polynomial, i.e., the value $a$ in \re{3.23.1} has to be a negative integer. If Kummer's function is a polynomial, then the function $\psi$ belongs to the Hilbert space $\mc H_2$ and solves the equation \re{2.3.4.47.1} with $q=0$ and eigenvalue $\lam$. Hence, $\lam=\lam_i$, for some $i\in\N$,  and $\psi$ is a multiple of the corresponding eigenfunction $u_i$ which is derived by the variational process in \rt{3.8}, since the sequence $(u_i)$ is complete and the eigenvalues have multiplicity one.

Qualitatively, it makes no difference, if $q=0$ or if $q\not=0$ satisfying
\begin{equation}
-2<q<2
\end{equation}
 is chosen. The plot will always be very similar to the graphics below which shows a Mathematica generated plot, where the parameters $m, m2, a$ correspond to the parameters in \re{2.3.4.62.1} with the exceptions that  $\mu_\e$ has been replaced by $m$ and $m_2$ by $m2$. Let us also emphasize that $m_2^2$ is a multiple of $\abs\Lam$.  The name $HypergeometricU$ is Mathematica's notation for Kummer's function $M(a,b,z)$. The eigenfunction is automatically evenly mirrored to the negative axis.
 
\begin{center}
\includegraphics[width=0.9\textwidth]{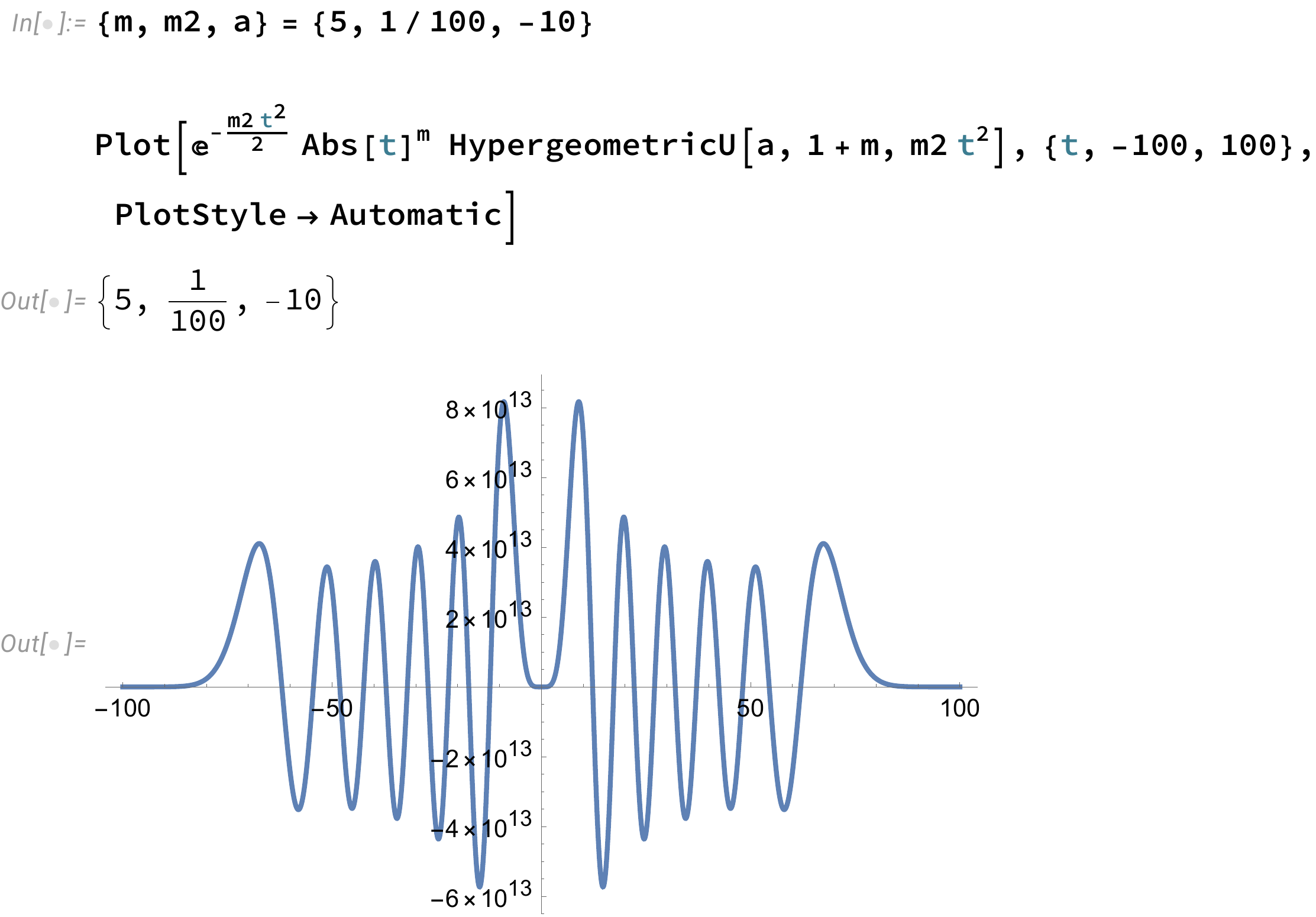}
\end{center}

Next let us prove sharp asymptotic estimates near infinity for the solutions of equation \re{3.111}.

\bl\lal{3.19} 
Let $\tilde u\in \tilde{\mc H}_2$ be a non-trivial solution of the equation \re{3.111}, where $\tilde\mu, m_2$ and $\lam$ are strictly positive and the exponent $q$ satisfies \re{3.2}. Then, for any $\e>0$ there exists $t_0>1$ and positive constants $c_1, c_2$ such that $\tilde u$ does not vanish in the interval $[t_0,\un)$ and can be estimated by
\begin{equation}
c_1 e^{-\frac12 (m_2+\e)t^2}\le \abs{\tilde u(t)}\le c_2 e^{-\frac12 (m_2-\e)t^2}\qq\A\, t\in [t_0,\un).
\end{equation}
\el
\bp
The proof is very similar to the proof \rt{2.3.4.10.1}. The first part, that $\tilde u$ does not vanish in the interval if $t_0$ is large enough, is almost identical. Suppose there would exist a large $t_0$ such that
\begin{equation}
\tilde u(t_0)=0
\end{equation}
and let us also consider a real valued $\tilde u$. Then, multiplying the equation \re{3.111} by $\tilde u$  and integrating by parts over the interval $[t_0,\un)$ would yield
\begin{equation}
\int_{t_0}^\un m_2^2 t^2 \tilde u^2\le \lam \int_{t_0}^\un t^q\tilde u^2,
\end{equation}
a contradiction if $t_0$ is large enough.

Thus, let us assume that $\tilde u$ is strictly positive in the interval and let us consider the comparison function
\begin{equation}
\psi(t)=e^{-\frac12 \mu t^2},
\end{equation}
where $\mu>0$ is an arbitrary constant. Then $\psi$ satisfies in $[t_0,\un)$
\begin{equation}
\begin{aligned}
&\qq-\Ddot \psi +\tilde\mu^2t^{-2}\psi+m_2^2t^2\psi-\lam t^q\psi\\
&=-\mu^2 t^2 \psi+\mu \psi+\tilde\mu^2t^{-2}\psi+m_2^2t^2\psi-\lam t^q\psi
\end{aligned}
\end{equation}
The left-hand side of the preceding equation defines a differential operator $A\psi$ and by an appropriate choice of $t_0$ and $\mu$ we deduce
\begin{equation}
A\psi=
\begin{cases}
<0,& \tup{if}\q m_2^2<\mu^2 ,\\
>0, & \tup{if}\q m_2^2 >\mu^2,
\end{cases}
\end{equation}
i.e., $\psi$ is a sub- \resp a super-solution depending on the choice of $\mu$ for the operator $A$ if $t_0$ is large enough. Next, by choosing $\e>0$ arbitrary but small we define
\begin{equation}
\mu=m_2+\e
\end{equation}
 and the constant $c_1>0$ such that
 \begin{equation}
\psi_1(t_0)\equiv c_1 e^{-\frac12 (m_2+\e)t_0^2}=u(t_0)
\end{equation}
to obtain a sub-solution $\psi_1(t)$, and similarly, a super-solution $\psi_2(t)$ is given by choosing
\begin{equation}
\psi_2(t_0)\equiv c_2 e^{-\frac12 (m_2-\e)t_0^2}=u(t_0).
\end{equation}
It is then fairly easy to infer
\begin{equation}\lae{3.133}
\psi_1(t)\le \tilde u(t)\le \psi_2(t)\qq\A\, t\in [t_0,\un).
\end{equation}
Indeed, to prove the second inequality we multiply the inequality
\begin{equation}
A(\tilde u-\psi_2)\le 0 
\end{equation}
in $[t_0,\un)$ by $\max(\tilde u-\psi_2,0)$ and integrating by parts to deduce
\begin{equation}
\int_{t_0}^\un (m_2^2 t^2-\lam t^q)\max(\tilde u-\psi_2,0)^2\le 0,
\end{equation}
which implies $\max(\tilde u-\psi_2,0)=0$  by the choice of $t_0$. Hence, we conclude
\begin{equation}
\tilde u\le \psi_2.
\end{equation}
The first inequality in \re{3.133} can be similarly derived, completing the proof of the lemma.
\ep

\bc\lac{3.20}
The unitarily equivalent eigenfunctions $w_i$ can be estimated near infinity by
\begin{equation}
c_1 t^{-\frac{m+k}2}e^{-\frac12 (m_2+\e)t^2}\le \abs{w_i(t)}\le c_2 t^{-\frac{m+k}2}e^{-\frac12 (m_2-\e)t^2}\;\A\, t\in [t_0,\un).
\end{equation}
\ec
\bp
Obvious, since
\begin{equation}
w_i(t)=t^{-\frac{m+k}2} \tilde u_i(t).
\end{equation}
\ep

\section{Trace class estimates for $e^{-\beta  H_0}$}    

We consider the operator $H_0$ in \fre{3.84} which  is essentially self-adjoint in
\begin{equation}
\mc H= L^2(\R[]_+,d\mu),
\end{equation}
where
\begin{equation} 
d\mu=\f_0dt
\end{equation}
with
\begin{equation}
\f_0(t)=t^{q},
\end{equation}
where $q$ satisfies the relation \re{3.87}
and we shall also use the same symbol for its closure, i.e., we shall assume that $H_0$ is self-adjoint in $\mc H$ with eigenvectors $u_i\in \tilde{\mc H}_2$  and with eigenvalues $\lam_i$ satisfying the statements in \frt{3.8}. However, now we denote the eigenvectors by $u_i$ to improve the readability. 

\br\lar{4.1}
The norm
\begin{equation}
\spd {H_0u}u^\frac12
\end{equation}
is equivalent to the norm $\nnorm u_2$ in $\tilde{\mc H}_2$, since $\abs\Lam>0$.

Let us also assume that all Hilbert spaces are complex vector spaces with a positive definite sesquilinear form (hermitian scalar product). 
\er

We shall now prove that 
\begin{equation}
e^{-\bet H_0},\qq\bet>0,
\end{equation}
is of trace class in $\mc H$. The proof is essentially the proof given in \cite[Chapter 3.5]{cg:qgravity-book2} with the necessary modifications due to the different exponent in $\f_0(t)$. 

First, we need two lemmata:  
\bl\lal{4.2}
The embedding
\begin{equation}
j:\tilde{\mc H}_2\hra \mc H_0=L^2(\R[]_+,d\tilde\mu),
\end{equation}
where
\begin{equation}
d\tilde\mu =(1+t)^{-2}dt, 
\end{equation}
is Hilbert-Schmidt, i.e., for any ONB $(e_i)$ in $\tilde{\mc H}_2$ the sum
\begin{equation}\lae{2.8.6.9}
\sum_{i=0}^\un\norm{j(e_i)}_0^2<\un
\end{equation}
is finite, where $\norm{\cdot}_0$ is the norm in $\mc H_0$. The square root of the left-hand side of \re{2.8.6.9} is known as the Hilbert-Schmidt norm  $\abs j$ of $j$ and  it is independent of the ONB, \cf \cite[Lemma 1, p. 158]{maurin:book}.    
\el

\bp
Let $w\in \tilde{\mc H}_2$, then, assuming $w$ is real valued, 
\begin{equation}\lae{2.8.6.2.23}
\begin{aligned}
\abs{w(t)}^2&=2\int_0^t\dot ww\le \int_o^\un \abs{\dot w}^2+\int_0^\un \abs{w}^2\\
&\le c \nnorm w_2^2
\end{aligned}
\end{equation}
for all $t>0$, where $\nnorm\cdot_2$ is the norm in $\tilde{\mc  H}_2$. To derive the last inequality in \re{2.8.6.2.23} we used \fre{3.33}.  The estimate  
\begin{equation}\lae{4.10}
\abs{w(t)}\le c\nnorm w_2\qq\A\, t>0
\end{equation}
is of course also valid for complex valued functions from which infer that, for any $t>0$, the linear form
\begin{equation}
w\ra w(t), \qq w\in \tilde{\mc H}_2,
\end{equation}
is continuous, hence it can be expressed as
\begin{equation}
w(t)=\spdd{\f_t}w_2,
\end{equation}
where
\begin{equation}
\f_t\in \tilde{\mc H}_2
\end{equation}
and
\begin{equation}
\nnorm{\f_t}_2\le c,
\end{equation}
in view of \re{4.10}. Now, let
\begin{equation}
e_i\in \tilde{\mc H}_2
\end{equation}
be an ONB, then
\begin{equation}
\begin{aligned}
\sum_{i=0}^\un\abs{e_i(t)}^2=\sum_{i=0}^\un\abs{\spdd{\f_t}{e_i}_2}^2=\nnorm{\f_t}_2^2\le c^2.
\end{aligned}
\end{equation}
Integrating this inequality over $\R[]_+$ with respect to $d\tilde\mu$ we infer
\begin{equation}
\sum_{i=0}^\un\int_0^\un\abs{e_i(t)}^2d\tilde\mu\le c^2
\end{equation}
completing the proof of the lemma.
\ep
\bl\lal{4.3}
Let $u_i$ be the eigenfunctions of $H_0$, then there exist positive constants $c$ and $\ga$ such that
\begin{equation}\lae{4.18}
\nnorm{u_i}_2\le c\abs{1+\lam_i}^\ga\norm{u_i}_0\qq\A\, i\in\N,
\end{equation}
where $\norm\cdot_0$ is the norm in $\mc H_0$. 
\el
 \bp
 We have
 \begin{equation}\lae{4.19}
\spd{H_0u_i}{u_i}=\lam_i\spd{u_i}{u_i}
\end{equation}
 and hence,  in view of \rr{4.1},
 \begin{equation}\lae{4.20}
\begin{aligned}
\nnorm{u_i}_2^2&\le c_1\lam_i\int_0^\un \f_0(t)\abs{u_i}^2\\
&\le c_1\lam_i\bigg\{\int_0^{1}\f_0(t)\abs{u_i}^2+c_2\int_{1}^\un t^{2-\frac2{l_0}}\abs{u_i}^2\bigg\},
\end{aligned}
\end{equation} 
where $l_0$ is very large such that
\begin{equation}
q<2-\frac2{l_0}.
\end{equation}
To estimate the second integral in the braces let us define $p=2$ and such that 
\begin{equation}
t^q\le t^{2-\frac2{l_0}}= t^{p-\frac p{l_0}}\qq\A\,t\ge 1.
\end{equation}
Then, choosing small positive constants $\de$ and $\e$, we apply Young's inequality, with
\begin{equation}
q_0=\frac p{p-p\de}=\frac 1{1-\de}
\end{equation}
and
\begin{equation}
q_0'=\de^{-1}
\end{equation}
to estimate the integral from above by
\begin{equation}
\begin{aligned}
\frac1{q_0}\e^{q_0}\int_{1}^\un\big\{t^{p-\frac p{l_0}}&(1+t)^{\frac p{l_0} -p\de}\big\}^{q_0}\abs{u_i}^2\\
&+\frac 1{q_0'}\e^{-q_0'}\int_{1}^\un (1+t)^{-(\frac p{l_0} -p\de)q_0'}\abs{u_i}^2.
\end{aligned}
\end{equation}
Choosing now $\de$ so small such that
\begin{equation}
(\frac p{l_0} -p\de)\de^{-1}>2
\end{equation}
the preceding integrals can be estimated from above by
\begin{equation}
\begin{aligned}
\frac 1{q_0}\e^{q_0}\int_{1}^\un (1+t)^p\abs{u_i}^2+\frac 1{q_0'} \e^{-q_0'}\int_0^\un (1+t)^{-2}\abs{u_i}^2
\end{aligned}
\end{equation}
which in turn can be estimated by
\begin{equation}
\frac 1{q_0} \e^{q_0} c \nnorm{u_i}_2^2+\frac 1{q'_0}\e^{-q_0'}\norm{u_i}_0^2,
\end{equation} 
in view of  \fre{3.33}.

Since $-2<q$ there exists $\e_0$ such that         
\begin{equation}
-(2-2\e_0)<q,
\end{equation}
hence, using again Young's inequality, the first integral in the braces on the right-hand side of \re{4.20} can be estimated by
\begin{equation}
\begin{aligned}
\int_0^{1}\f_0(t)\abs{u_i}^2&\le c\int_0^1t^{-(2-2\e_0)}\abs{u_i}^2\le c (1-\e_0) \e^{\frac1{1-\e_0}}\int_0^1t^{-2}\abs{u_i}^2\\
&\qq+c \e_0\e^{-\frac1{\e_0}}\int_0^\un (1+t)^{-2}\abs{u_i}^2\\
&\le \tilde c(1-\e_0) \e^{\frac1{1-\e_0}}\nnorm{u_i}_2^2+c \e_0\e^{-\frac1{\e_0}}\norm{u_i}_0^2.
\end{aligned}
\end{equation}
Choosing now $\e,\ga$ and $c$ appropriately the result follows.
 \ep
We are now ready to prove:
\bt\lat{2.8.6.4}
Let $\bet>0$, then the operator
\begin{equation}
e^{-\bet H_0}
\end{equation}
is of trace class in $\mc H$, i.e.,
\begin{equation}
\tr(e^{-\bet H_0})=\sum_{i=0}^\un e^{-\bet\lam_i}=c(\bet)<\un.
\end{equation}
\et
\bp
In view of \rl{4.2} the embedding 
\begin{equation}
j:\tilde{\mc H}_2\hra\mc H_0
\end{equation}
is Hilbert-Schmidt. Let 
\begin{equation}
u_i\in \mc H
\end{equation}
be an ONB of eigenfunctions, then
\begin{equation}
\begin{aligned}
e^{-\bet\lam_i}&=e^{-\bet\lam_i}\norm{u_i}^2\le e^{-\bet\lam_i}c\lam_i^{-1}\nnorm{u_i}_2^2\\
&\le e^{-\bet\lam_i}\lam_i^{-1}c\abs{\lam_i+1}^{2\ga}\norm{u_i}_0^2,
\end{aligned}
\end{equation}
in view of \re{4.19} and \re{4.18}, but
\begin{equation}
\begin{aligned}
\norm{u_i}_0^2=\nnorm{u_i}_2^2\,\norm{\tilde u_i}_0^2\le c\lam_i\norm{\tilde u_i}_0^2,
\end{aligned}
\end{equation}
where
\begin{equation}
\tilde u_i=u_i \nnorm{u_i}_2^{-1}
\end{equation}
is an ONB in $\mc H_2$, yielding
\begin{equation}
\sum_{i=0}^\un e^{-\bet\lam_i}\le  c_\bet \sum_{i=0}^\un\norm{\tilde u_i}_0^2<\un,
\end{equation}
since $j$ is Hilbert-Schmidt. Here we  used  \rr{4.1}, since the scalar product in $\tilde {\mc H}_2$ has to be defined by
\begin{equation}
\spd{H_0 u}{v}
\end{equation}
in order to deduce that the eigenfunctions are also mutually orthogonal in $\tilde{\mc H}_2$,
 and  also $\lam_0>0$.
\ep

\br
This result enables us to apply quantum statistics to our model of quantum gravity and to define
a partition function $Z$, a density operator $\rho$ and the von Neumann entropy $S$ in a corresponding Fock space. For details we refer to \cite[Chapter 9.5]{cg:qgravity-book2}.
\er

\section{Conclusions}\las{5}
In our model of quantum gravity the physical states are described by solutions of a hyperbolic equation in a fiber bundle with base space $\socc$ which is isometric to a Cauchy hypersurface of the quantized spacetime. The solutions of the hyperbolic equation can be expressed as a product of temporal and spatial eigenfunctions of self-adjoint operators acting in appropriate Hilbert spaces. The coefficients of the  temporal eigenfunction equation as well as the corresponding eigenfunctions $w_i$ have a singularity in $t=0$ similar to the big bang singularity of the quantized spacetime.  

However, by introducing a scalar field map 
\begin{equation}
\F: \socc \ra \R[k]
\end{equation}
 in the quantization process we proved in \frt{3.15} that there exist a complete sequence of unitarily equivalent temporal eigenfunctions $\tilde u_i$ which solve the eigenfunction equation
\begin{equation}\lae{5.2}
-\Ddot{ \tilde u}_i  + t^{-2} {\tilde \mu}^2 \tilde u _i+t^2 m_2^2  \tilde u_i=\lam_i \abs t^q  \tilde u,
\end{equation}
in the interval $(0,\un)$, where $\tilde\mu, m_2$ and $\lam_i$ are strictly positive and
\begin{equation}
-2<q<2
\end{equation}
is a fixed exponent, such that the solutions $\tilde u_i$ can be evenly or oddly mirrored to the negative axis  as sufficiently smooth functions across the singularity provided the dimension $k$ of the target space of $\F$ is sufficiently large. The equation \re{5.2} is then valid in $\R[]$ such that both sides smoothly vanish in $t=0$.
A plot before \frl{3.19} illustrates the situation. Moreover, we proved in \rl{3.19} that the eigenfunctions $\tilde u_i$ vanish exponentially fast near infinity which is also valid for the unitarily equivalent eigenfunctions $w_i$, \cf \frc{3.20}. 

The hyperbolic equation in the fiber bundle comprised second order differential operators acting in the fibers as well as in the base space. The temporal equation, we consider here,  also defines a second order differential operator acting in the fibers because the Riemannian metrics, which are part of the variables after quantization, can be written in the form
\begin{equation}
g_{ij}=t^\frac 4n \s_{ij},
\end{equation}
where $0<t<\un$ and the $\s_{ij}(x)$, $x\in \socc$, are elements of a subbundle with fibers $M(x)$ such that by fixing an arbitrary metric $\bar\s_{ij}(x)$ which is supposed to be the induced metric of a Cauchy hypersurface of the quantized spacetime, we may assume, after choosing an appropriate atlas depending on $\bar\s_{ij}$, that each fiber $M(x)$  is isometric to the symmetric space
\begin{equation}
X=SL(n,\R[])/SO(n),
\end{equation}
\cf  \rs{1}. The elementary gravitons are then  eigenfunctions of the Laplacian in $X$. The corresponding eigenvalues are already incorporated in the coefficient $\tilde\mu^2$ of the temporal differential operator such that we are allowed to consider, besides the temporal operator, only  spatial operators acting in $\socc$.

Thus, we look at a quantum spacetime $Q$ which can be written as a product
\begin{equation}
Q=(0,\un)\times \socc
\end{equation}
and at self-adjoint operators $H_0$ and $H_1$ acting in appropriate Hilbert spaces such that the remaining hyperbolic equation in $Q$ can be expressed in the form
\begin{equation}\lae{5.7}
H_0u-H_1u=0,
\end{equation}
where $u$ is a product of temporal and spatial eigenfunctions of $H_0$ \resp $H_1$
\begin{equation}
u(t,x)=\tilde u_i(t) \psi_j(x)
\end{equation}
\cf \cite{cg:uqtheory3b,cg:qgravity4,cg:qgravity-book2} for more details.  

Since the temporal eigenfunctions can be smoothly mirrored to the negative axis we may consider a second quantum spacetime
\begin{equation}
Q_{-}=(-\un, 0)\times \socc
\end{equation}
in which the equation \re{5.7} is also valid. Moreover, the equation \re{5.2} is even valid in $\R[]$ across  the singularity. Hence, we have to face the question how to interpret this behaviour. If  we assume that $Q_{-}$ has the same light cone as $Q$, then the singularity in $t=0$ lies in the future of $Q_-$ and since the mirrored eigenfunctions $w_i(t)$ become unbounded if $t$ tends to zero,  the singularity in $t=0$ would be called a big crunch, i.e., $Q_-$ would end in a big crunch but the corresponding classical spacetime would not start with a big bang in view of the results in \rl{3.19} and \frc{3.20}.

Hence, we have to assume that $Q_-$ has the opposite time orientation, i.e., the singularity in $t=0$ is also a big bang for $Q_-$. In \cite{cg:uqtheory3b} and \cite[Chapter 5]{cg:qgravity-book2} we proved that we may consider $H_1$ to be a spatial self-adjoint operator defined by the fields of the Standard Model. If $H_1$ is invariant with respect to parity and charge conjugation then, in view of the CPT theorem,  we would conclude that at the big bang two universes had been created with different time orientation one filled with matter and the other with antimatter.

\bibliographystyle{hamsplain}

\begin{thebibliography}{10}

\bibitem{adm:old}
R.~Arnowitt, S.~Deser, and C.~W. Misner, \emph{The dynamics of general
  relativity}, Gravitation: an introduction to current research (Louis Witten,
  ed.), John Wiley, New York, 1962, pp.~227--265.

\bibitem{courant-hilbert-I}
R.~Courant and D.~Hilbert, \emph{Methoden der mathematischen {P}hysik. {I}},
  Springer-Verlag, Berlin, 1968,
  {\href{https://doi.org/10.1007/978-3-642-47436-1}{doi:10.1007/978-3-642-47436-1}},
  Dritte Auflage, Heidelberger Taschenb\"ucher, Band 30.

\bibitem{cg:pdeII}
Claus Gerhardt, \emph{{Partial differential equations II}}, Lecture Notes,
  University of Heidelberg, 2013,
  {\href{http://www.math.uni-heidelberg.de/studinfo/gerhardt/PDE2.pdf}{pdf
  file}}.

\bibitem{cg:qgravity}
\bysame, \emph{{The quantization of gravity in globally hyperbolic
  spacetimes}}, Adv. Theor. Math. Phys. \textbf{17} (2013), no.~6, 1357--1391,
  {\href{http://arXiv.org/abs/1205.1427}{arXiv:1205.1427}},
  {\href{http://dx.doi.org/10.4310/ATMP.2013.v17.n6.a5}{doi:10.4310/ATMP.2013.v17.n6.a5}}.

\bibitem{cg:qgravity3}
\bysame, \emph{{The quantization of gravity: Quantization of the Hamilton
  equations}}, Universe \textbf{7} (2021), no.~4, 91,
  {\href{http://dx.doi.org/10.3390/universe7040091}{doi:10.3390/universe7040091}}.

\bibitem{cg:uqtheory3b}
\bysame, \emph{{A unified quantization of gravity and other fundamental forces
  of nature}}, Universe \textbf{8} (2022), no.~8, 404,
  {\href{https://doi.org/10.3390/universe8080404}{doi:10.3390/universe8080404}}.

\bibitem{cg:qgravity4}
\bysame, \emph{{The quantization of gravity: The quantization of the full
  Einstein equations}}, Symmetry \textbf{15} (2023), no.~8, 1599,
  {\href{https://doi.org/10.3390/sym15081599}{doi:10.3390/sym15081599}}.

\bibitem{cg:qgravity-book2}
\bysame, \emph{{The Quantization of Gravity}}, 2nd ed., Fundamental Theories of
  Physics, vol. 194, Springer, Cham, November 2024,
  {\href{https://dx.doi.org/10.1007/978-3-031-67922-3}{doi:10.1007/978-3-031-67922-3}}.

\bibitem{kamke:ode}
Erich Kamke, \emph{{Gew{\"o}hnliche Differentialgleichungen}}, {10.Aufl.,
  unver{\"a}nd. Nachdr. d. 8., durchges. Aufl.} ed., Teubner, Stuttgart, 1983
  (ger).

\bibitem{maurin:book}
Krzysztof Maurin, \emph{Methods of {H}ilbert spaces}, Translated from the
  Polish by Andrzej Alexiewicz and Waclaw Zawadowski. Monografie Matematyczne,
  Tom 45, Pa\'nstwowe Wydawnictwo Naukowe, Warsaw, 1967.

\bibitem{nist:handbook}
Frank W.~J. Olver, Daniel~W. Lozier, Ronald~F. Boisvert, and Charles~W. Clark
  (eds.), \emph{{NIST Handbook of Mathematical Functions}}, Cambridge
  University Press, Cambridge; New York; Melbourne, 2010 (eng).

\end{thebibliography}
\providecommand{\bysame}{\leavevmode\hbox to3em{\hrulefill}\thinspace}
\providecommand{\href}[2]{#2}



\end{document}